\documentclass[fleqn,usenatbib]{mnras}

\usepackage{newtxtext,newtxmath}
\usepackage{ulem}
\usepackage[T1]{fontenc}

\DeclareRobustCommand{\VAN}[3]{#2}
\let\VANthebibliography\thebibliography
\def\thebibliography{\DeclareRobustCommand{\VAN}[3]{##3}\VANthebibliography}

\usepackage{graphicx}	% Including figure files
\usepackage{amsmath}	% Advanced maths commands
\title[Lensing of standard sirens]{Lensing bias on cosmological parameters from bright standard sirens}

\author[S. Canevarolo et al.]{
Sofia Canevarolo,$^{1}$\thanks{E-mail: s.canevarolo@uu.nl}
Nora Elisa Chisari$^{1}$\thanks{E-mail: n.e.chisari@uu.nl}
\\
$^{1}$ Institute for Theoretical Physics, Utrecht University, Princetonplein 5, 3584 CC, Utrecht, The Netherlands.
}

\date{Accepted XXX. Received YYY; in original form ZZZ}

\pubyear{2023}

\begin{document}
\label{firstpage}
\pagerange{\pageref{firstpage}--\pageref{lastpage}}
\maketitle

\begin{abstract}
Next generation gravitational waves (GWs) observatories are expected to measure GW signals with unprecedented sensitivity, opening new, independent avenues to learn about our Universe. The distance-redshift relation is a fulcrum for cosmology and can be tested with GWs emitted by merging binaries of compact objects, called standard sirens, thanks to the fact that they provide the absolute distance from the source. On the other hand, fluctuations of the intervening matter density field induce modifications on the measurement of luminosity distance compared to that of a homogeneous universe. Assuming that the redshift information is obtained through the detection of an electromagnetic counterpart, we investigate the impact that lensing of GWs might have in the inference of cosmological parameters. We treat lensing as a systematic error and check for residual bias on the values of the cosmological parameters.  We do so by means of mock catalogues of bright sirens events in different scenarios relevant to Einstein Telescope. For our fiducial scenario, the lensing bias can be comparable to or greater than the expected statistical uncertainty of the cosmological parameters, although non-negligible fluctuations in the bias values are observed for different realisations of the mock catalogue. We also discuss some mitigation strategies that can be adopted in the data analysis. Overall, our work highlights the need to model lensing effects when using standard sirens as probes of the distance-redshift relation.
\end{abstract}


%%%%%%%%%%%%%%%%%%%%%%%%%%%%%%%%%%%%%%%%%%%%%%%%%%

%%%%%%%%%%%%%%%%% BODY OF PAPER %%%%%%%%%%%%%%%%%%

\section{Introduction}

Since the first detection in 2015 by the LIGO-Virgo collaboration \citep{LIGO2015}, gravitational waves (GWs) have opened a new observational window on our Universe. As first observed by \citet{Schutz1986}, the gravitational wave signal from the inspiral and merger of a compact binary system carries information about our distance from the source. In fact, the amplitude of the GW signal from compact binary systems is inversely proportional to the luminosity distance of its source. This useful property turns merging binaries emitting GWs into \textit{standard sirens} \citep{Holz2005, Dalal2006}, suitable for probing the expansion history of the Universe. This is of crucial importance to solve many open problems in cosmology, such as the current tension on the value of the Hubble constant $H_0$ in the $\Lambda \rm CDM$ model and to investigate possible extensions of the cosmological paradigm \citep[see][and references therein]{DiValentino2021, Perivolaropoulos2022}.

To probe cosmology with standard sirens, redshift information is also necessary. However, in the waveform of the standard sirens, the redshift is degenerate with a combination of the masses of the two objects in the binary system - the so-called mass-redshift degeneracy. The detection of an electromagnetic (EM) counterpart of the GW signal can provide an independent measurement of the redshift, breaking the degeneracy. A strategy for the detection of the EM counterpart can consist of an optical/IR follow-up in the localization area delimited by the GW detectors, aimed at the identification of the host galaxy, whose redshift can be measured spectroscopically or photometrically. If the localization area of the GW detectors is very broad, or even in the case that no localization region is provided, it is possible to exploit the coincident detection of a gamma ray burst \citep[GRB][]{Dalal2006}, and the corresponding $X$-ray/optical afterglow emission. In particular, the afterglow is important for the host galaxy identification.

In August 2017, the first coincident detection of the GW event GW170817 from the coalescence of two neutron stars and its EM counterpart was achieved \citep{Abbott_2017_GRB}, opening the era of multi-messenger cosmology. The EM counterpart, the GRB event, GRB170817A, followed by an afterglow in other frequencies of the EM spectrum, allowed the localization of the host galaxy of the source and led to the first constraint on the Hubble constant by GWs \citep{Abbott2017}. At present, the bound obtained with the Third Gravitational-Wave
Transient Catalogue \citep{Abbott2021} is $H_0=69^{+12}_{-7}\,\rm km\,s^{-1}Mpc^{-1}$ \citep{Gray2023}. These measurements, even if not yet comparable with the cosmological constraints already achieved by other EM probes, give proof that the distance-redshift relation, and thus cosmological parameters, can be independently tested with standard siren events. 

In a decade from now, a third generation (3G) of ground- and space-based GW detectors, such as the Einstein Telescope (ET) \citep{Punturo2010, Branchesi2023}, Cosmic Explorer (CE) \citep{Reitze2019, Evans2021} and LISA \citep{Amaro-Seoane2017}, are expected to measure GW signals with unprecedented sensitivity, reaching very high redshift events and accumulating an enormous number of detections. In this paper, we focus on the signals coming from Neutron Star binary (BNS) coalescence, which are the best candidate events to be accompanied by an EM counterpart, the so called bright sirens. The Einstein Telescope is expected to detect $\sim 10^5$ BNS per year up to redshifts of $z \simeq 2-3$ \citep{ETscience}. A fraction of these events will be bright sirens and will allow for precise GW cosmology \citep{Sathyaprakash2010}. By the time ET enters its full operational regime, new EM facilities, such as {\it Euclid} \citep{Laureijs2011}, the Vera C. Rubin Observatory \citep{Ivezi2019} and DESI \citep{Hahn2023}, will have provided spectroscopic and photometric galaxy catalogues useful for the host galaxy determination.

Similarly to the case of light, GW signals are expected to be affected by gravitational lensing. If the alignment between the source, the lens and the observer is favorable, the gravitational waves can be strongly lensed, which can result in multiple images of the same signal. Searches for strongly lensed events have been carried out in LIGO-Virgo data but it is still a matter of discussion whether these are affected by lensing \citep{Abbott2021_lensing,Diego2021}. The observation of such events would allow a plethora of fundamental and cosmological applications, widely explored in the literature \citep[some examples are][]{Sereno2011,Baker2017,Liao2017,Goyal2021}. Strongly lensed events are expected to constitute a small fraction of the huge detection rates of the 3G detectors \citep{Oguri2018,Li2018}. On the contrary, weak lensing caused by the gravitational potential of the large scale structures in the Universe is ubiquitous, and accumulates along the path of GWs, thus becoming more relevant for high redshift sources \citep{Holz2005}. 

In this paper, we investigate the impact of lensing on the  estimation of cosmological parameters from bright standard sirens measurements. We treat both weak and strong scenarios simultaneously based on the statistical distribution of values for the lensing magnification, even when it is very high. The prospect of the lensing of GWs is two-fold. It can be regarded as an extra source of information enabling the use of GWs to study the matter distribution along the line-of-sight, thus testing the standard model of cosmology and shedding light on the nature of dark energy \citep{Congedo2019, Mukherjee2020,Congedo2022, Balaudo2023}. Or, it can be regarded as a nuisance in the determination of the distance-redshift relation with the standard sirens method. In such a case, it is normally accounted for as a source of noise in the distance measurement \citep{Sathyaprakash2010,Hirata2010}. Here we consider the latter view, but we treat lensing as a {\it systematic error} and we check whether the statistical uncertainty on the luminosity distance fully absorbs any bias of the cosmological parameters.

Two distinct aspects can source this bias: lensing selection effect and the biased estimation of the luminosity distance from the GW signal. The former effect, already studied in \citet{Cusin2019, CusinTamanini2021,Shan2021}, is due to the fact that we are more likely to detect magnified events, whose Signal to Noise Ratio (SNR) is increased by the lensing magnification factor \citep{Dai2017}. In fact, lensing can act as a magnifying glass, enabling the detection of faint and far events, not accessible in a homogeneous universe. On the other hand, the non-Gaussian shape of the lensing magnification probability distribution function (PDF) might also yield a bias in the estimation of the luminosity distance. As we will see, the lensing PDF is skewed, with a tail towards high magnifications, produced by the rare overdensities of the matter field in the Universe. This highlights the importance of investigating lensing bias in the context of GW cosmology.

We adopt a flat FLRW cosmology with the following fiducial values of the parameters: $h=H_0/100\, \rm km \,s^{-1}Mpc^{-1}=0.6736$, $\Omega_{\rm b} \,h^2=0.02237$, $\Omega_{\rm c}\, h^2=0.12$, where $\Omega_{\rm b}$ and $\Omega_{\rm c}$ are respectively the baryon and cold dark matter density parameters. These values correspond to those adopted in the AbacusSummit simulations of \citet{Hadzhiyska2023}, which we use to model gravitational lensing. 

The paper is organised as follow. In Section \ref{sec:theory}, we review the theory of bright standard sirens and the impact of lensing. In Section \ref{sec:Mock}, we build a realistic mock catalogue of lensed bright sirens events, detectable by a third generation ground-based detector, i.e. ET. We also compare the lensing magnification PDF obtained from lensing convergence maps with a simple log-normal analytical model, often used in the literature \citep{Kayo2001,Das2006, Martinelli2022}. In Section \ref{sec:Method}, we present the methods adopted to estimate the bias on the cosmological parameters, which exploit the Fisher matrix formalism and the full Monte Carlo Markov Chain (MCMC) analysis. The results of our study are then presented in Section \ref{sec:Results}, where different scenarios are considered. In Section \ref{sec:Mitigation}, we present a procedure to slightly mitigate the effect of lensing for the cosmological parameter estimation analysis. Finally, in Section~\ref{sec:Discussion}, we present our discussion of the results and in Section~\ref{sec:Conclusion} the conclusions.

\section{Lensed standard sirens}
\label{sec:theory}
At leading order, in frequency domain and using the stationary phase approximation, the waveform of the inspiral of two compact objects can be written as \citep{Cutler1994},
\begin{equation}
    \Tilde{h}(f)= \sqrt{\frac{5}{96}} \frac{\mathcal{Q}}{d_{\rm L}}\frac{(G \mathcal{M}(1+z))^{5/6}}{\pi^{2/3}c^{3/2}}f^{-7/6} e^{-i\Phi(f,\mathcal{M}(1+z))},
    \label{waveform}
\end{equation}
with
\begin{equation}
    \mathcal{Q}=\sqrt{F_+^2(\theta,\phi,\psi)(1+\cos^2\iota)^2+F_{\times}^2(\theta,\phi,\psi)4\cos^2\iota}.
\end{equation}
$\mathcal{M}=\frac{(m_1 m_2)^{3/5}}{(m_1+m_2)^{1/5}}$ is the chirp mass of the binary composed by two individual objects of masses $m_1$ and $m_2$, $\Phi$ the phase of the strain, $G$ the Newton constant and $c$ the speed of light. The function $\mathcal{Q}$ depends on the position of the binary in the sky, determined by the angles $(\theta,\phi)$, the inclination angle $\iota$ and the polarization angle $\psi$. $F_{+,\times}$ are antenna pattern functions that describe the response of the detector.

The luminosity distance of the source, $d_{\rm L}$, is inversely proportional to the amplitude of the strain and is degenerate with the angles in the amplitude of the strain, in particular with the inclination angle $\iota$ \citep{Nissanke2010}. In a flat FLRW spacetime, $d_{\rm L}$ is given by:
\begin{equation}
    d_{\rm L}(z)=\frac{c\, (1+z)}{H_0}\int_0^z\frac{dz}{\sqrt{\Omega_{\rm m}(1+z)^3+\Omega_{\rm DE}(1+z)^{3(1+w_{\rm DE})}}},
    \label{d_l}
\end{equation}
where $\Omega_{\rm m}$ is the energy density in matter (baryons plus dark matter), $\Omega_{\rm DE}$ is the energy density in dark energy, and $w_{\rm DE}$ is the dark energy equation of state parameter. In a flat universe, $\Omega_{\rm DE}=1-\Omega_{\rm m}$. For a cosmological constant, $w_{\rm DE}=-1$.

In equation \eqref{waveform}, the redshift of the source $z$ is degenerate with the chirp mass, such that only the redshifted chirp mass, $\mathcal{M}_z=(1+z)\mathcal{M}$, can be measured from a GW signal. Additional information is needed to break this degeneracy. In this paper we assume that the redshift can be determined from the detection of an EM counterpart that allows the localization of the host galaxy of the compact binary coalescence. If the counterpart is a strongly beamed GRB emission, with $ \iota \lesssim 20 \degr$, \citet{Nissanke2010} pointed out that it can provide a prior on the inclination angle which helps break the $d_{\rm L} - \iota$ degeneracy. Alternatively, the measurement of higher order modes of the GW signals can also alleviate this degeneracy \citep{Bustillo2020}.

Possible GW sources that can be accompanied by an EM counterpart are BNS, a neutron star and a black hole (BHNS) and supermassive black holes (SMBH). SMBHs are expected to be detectable in the low frequency band covered by space-based detectors, such as LISA. In this paper, we choose the framework of ground-based detectors, in particular ET and we concentrate on the population of BNSs. We do not consider BHNS mergers, as their multi-messenger rate is quite uncertain \citep{Fragione2021}. Nevertheless, since the lensing universally affects all GW signals, we expect our methodology could be extended to other GW sources, in other frequency bands, with or without an EM counterpart.

\subsection{Lensing magnification}
As studied in \citet{Laguna2010,Bertacca2018, Fonseca2023}, different modifications of the GW waveform are expected due to the fluctuations of the intervening matter density field, which induce effects such as lensing, Doppler, Integrated Sachs-Wolfe, Sachs-Wolfe, Shapiro time-delays and volume distortions. In this paper, we only model the effect of lensing on the luminosity distance, which is expected to be dominant for $z \gtrsim 0.5$.
The geometric optics approximation, valid when the wavelength of the GW is much smaller than the size of the lens, predicts modifications that are frequency independent, i.e. the frequency of the lensed waves remain unchanged. Within this approximation, we expect the strain of the gravitational waves to be magnified or demagnified by the structures or voids the GWs encounter along their propagating path:

\begin{equation}
    h(f)=\sqrt{\mu}\, h_{\rm unlensed}(f),
\end{equation}
where $h_{\rm unlensed}(f)$ is the waveform computed in the homogeneous universe, e.g. it satisfies equation \eqref{waveform}, and we dropped the tilde that signals the frequency domain for simplicity. This modification translates into an observed luminosity distance $d_{\rm L}^{\rm bias}$ which differs with respect to the true one by \citep{Kocsis2006}:
\begin{equation}
    d_{\rm L}^{\rm bias}(z,\mu)=\frac{d_{\rm L}(z)}{\sqrt{\mu}},
    \label{dl_bias}
\end{equation}
where $\mu$ is the lensing magnification. At lowest order, the magnification is related to the lensing convergence field $\kappa$ by $\mu \simeq 1+2\kappa$. But more generally, the relation between convergence and magnification can be more accurately quantified by \citep[Figure 8]{Takahashi2011},
\begin{equation}
    \mu=\frac{1}{(1-\kappa)^2},
    \label{mu-kappa}
\end{equation}
which is derived from the definition of magnification, neglecting the effect of the shear. In this work, we use equation \eqref{mu-kappa} to allow for arbitrarily high values of $\mu$.

It is interesting to note that there exists another regime of  extreme magnification caused by the presence of microlenses along the line-of-sight. We do not attempt to model this regime explicitly in this work since we assume they could be separately identified from wave optics effects \citep{Takahashi2003, Diego2019, Mishra2021, Kim2022}.

\section{Mock catalog}
\label{sec:Mock}

\begin{figure}
\centering    
\includegraphics[width=0.45\textwidth]{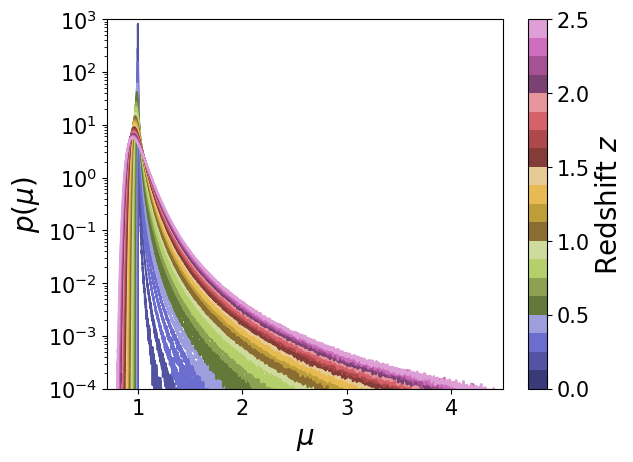}
\caption{Lensing magnification PDF for different source redshift, as obtained from the convergence maps of \citet{Hadzhiyska2023}.}
\label{fig:pmu}
\end{figure}

Assuming the redshift is known from the observation of an EM counterpart, we build a mock catalogue of BNS events following the steps: 
\begin{enumerate}
    \item We draw the redshift of the events from a distribution built assuming that the BNS merger rate tracks the Madau-Fragos star formation rate \citep{Madau2017}, and accounting for a time delay $t_{\rm d}$ between the redshift of formation $z_{\rm f}$ and of merger $z_{\rm m}$ \citep{Regimbau2012,Muttoni2023}.  The star formation rate is given by:
    \begin{equation}
        \Psi(z)=0.01\frac{(1+z)^{2.6}}{1+[(1+z)/3.2]^{6.2}} \;\; [\rm M_{\sun} Gpc^{-3} yr^{-1}],   
    \end{equation}
    and we assume the following probability distribution for the time delay,
    \begin{equation}
        p(t_{\rm d}) \propto \frac{1}{t_{\rm d}}.
    \end{equation}
    The theoretical merger rate is then:
    \begin{equation}
    \begin{split}
        p^{\rm (th)}(z)=\frac{1}{1+z}\frac{4 \pi c \chi^2(z)}{H(z)}\int_{t_{\rm d}^{\rm MIN}}^{t_{\rm d}^{\rm MAX}}& dt_{\rm d} \Psi(z_{\rm f}(z,t_{\rm d})) \\
        & \times \bigg( \log\bigg(\frac{t_{\rm d}^{\rm MAX}}{t_{\rm d}^{\rm MIN}}\bigg)t_{\rm d}\bigg)^{-1},
    \end{split}
    \end{equation}
    where $\chi(z)=c \int_0^z \frac{dz}{H(z)}$ is the comoving distance and $H(z)=H_0\sqrt{\Omega_{\rm m}(1+z)^3+\Omega_{\rm DE}}$ is the Hubble parameter.
    We assume a minimum time delay between formation and merger $t_{\rm d}^{\rm MIN}=20\; \rm Myr$ and a maximum $t_{\rm d}^{\rm MAX}=10\; \rm Gyr$ \citep{Borhanian2022}.
    
    \item Following \citet{Congedo2022}, we account for selection effects due to the coverage of galaxy surveys. We obtain the observed probability distribution for the source redshift as:
    \begin{equation}
        p(z)=p^{\rm (th)}(z)f(z)
    \end{equation}
    where $f(z)$ is the selection function of the galaxy survey and takes the form,
    \begin{equation}
        f(z)=\frac{0.9}{2}\bigg(1-\tanh\bigg(\frac{z-z_{\rm pivot}}{w z_{\rm pivot}}\bigg)\bigg), \;\;\; w=\frac{z_{\rm max}-z_{\rm pivot}}{2 z_{\rm max}},
        \label{EMsel}
    \end{equation}
    with $z_{\rm pivot}=1.8$ and $z_{\rm max}=2$. This is based on the maximum redshift coverage that future spectroscopic galaxy surveys are expected to reach \citep{Laureijs2011}. Beyond this limit, we assume that the redhsift of the source cannot be estimated via spectroscopy.
    
    We draw a set of $\{\hat{z}\}$ from the distribution $p(z)$.
    In Section~\ref{sec:Results}, we will also study the dependence of our results on the galaxy survey selection function.

    \item  We assign to each event a value of the lensing magnification. Specifically, for each redshift $\hat{z}_i$ in the collection $\{\hat{z}\}$, we assign a value $\hat{\mu}_i$, by drawing from the magnification PDF at redshift $\hat{z}_i$. 
    
    More details about the magnification PDF are given in subsection \ref{subsec:mu PDF}. For our base analysis, we consider the AbacusSummit suite of cosmological N-body simulations, presented in \citet{Hadzhiyska2023}, and the publicly available attached lensing distributions\footnote{\label{foot:AbacusSummit}\url{https://app.globus.org/file-manager?origin_id=3dd6566c-eed2-11ed-ba43-09d6a6f08166&path=\%2F}}. We use the ``base'' box with a pixel resolution of $0.21 \,\rm arcmin$ and a coverage of an octant of the sky, up to $z\sim 0.8$, then reduced to $1800\, \rm deg^2$. We obtain the magnification PDF from the maps of the convergence fields at all source redshifts, 47 in total, ranging from $z=0.15$ to $z=2.45$, and applying eq.~\eqref{mu-kappa}. The result is shown in Fig.~\ref{fig:pmu}. As already said, we do not attempt to distinguish between weak and strong lensing regime, and we allow for a tail of high magnifications, for example due to unidentified strongly lensed images. This can be the consequence of a single image event or of multiple images that arrive at the detector with long time delays and are recorded as separate events.
    \item We draw values for the masses of the neutron stars from a redshift independent distribution with a Gaussian shape centered in $\Bar{m}=1.35\, \rm M_{\sun}$ and with standard deviation $\sigma=0.15 \,\rm M_{\sun}$, normalised in the interval $[1;2]\rm M_{\sun}$ \citep{Borhanian2022}. We also consider the case of a uniform distribution between $[1;2.5]\rm M_{\sun}$ \citep{Belgacem2019,Ronchini2022}.
    
    \item We draw the angles $\theta$ and $\iota$ from isotropic distributions in the range $[0,\,\pi]$. The polarisation angle $\psi$ is drawn from a uniform distribution in the range $[0,\,\pi]$. If we assume that the EM counterpart is a beamed GRB, it might be possible to constrain the value of the inclination angle $\iota$. Therefore, we also consider the case in which we restrict $ \iota < 20 \degr$. 
    
    \item Using the mock collection of $\{\hat{z},\hat{\mu}\}$ and our fiducial values of the cosmological parameters we compute the luminosity distance of each event. We assume to infer the observed luminosity distance from the full parameters estimation analysis over all the extrinsic and intrinsic parameters that characterise the GW signal. The luminosity distance obtained from this analysis is biased by the effect of lensing, as given in equation \eqref{dl_bias}. 
    
    \item For each event in our theoretical mock catalog, we calculate the SNR $\rho$, defined by,
    \begin{equation}
        \rho^2=4 \int df \frac{|h(f)|^2}{S_{\rm n}(f)},
    \end{equation}
    where $S_{\rm n}(f)$ is the one-sided noise power spectral density (PSD) of a detector. In the case of ET, we have to sum over the three detectors that compose the triangular configuration. We assume the same PSD function, the ET-D sensitivity curve \footnote{\label{foot:ETwebsite}\url{https://www.et-gw.eu}}, for all the three detectors. In absence of correlated noise, the total SNR is:
    \begin{equation}
       \rho_{\rm tot}^2=\sum_{a=1}^3 \rho_a^2,
    \end{equation}
    where $\rho_a^2$ is the SNR of the individual detector. A realistic GW interferometer is sensitive to SNRs above $\rho_{\rm lim}$. We assume $\rho_{\rm lim}=12$, summed over the three independent detectors of ET \citep{Branchesi2023}, and we create a mock catalogue of detectable events above such threshold.
 
    For BNS events, which involve low mass objects, we can use the following approximated formula, valid during the inspiral part of the coalescence, to compute the SNR of the event \citep{Finn1993}:
    \begin{equation}
        \rho_{\rm tot}^2=\mu \frac{5}{6} \frac{(G\mathcal{M}(1+z))^{5/3}}{c^3 \pi^{4/3}d_{\rm L}^2(z)}\mathcal{F}^2(\theta,\iota,\psi)\int_{f_{\rm min}}^{f_{\rm max}} df \frac{f^{-7/3}}{S_{\rm n}(f)},
    \label{SNReq}
    \end{equation}
    with \citep{Regimbau2012}:
    \begin{equation}
    \begin{split}
        \mathcal{F}^2(\theta,\iota,\psi)=&\frac{9}{128}(1+\cos^2\iota)^2(1+\cos^2\theta)^2\cos^22\psi\\
        &+\frac{9}{32}(1+\cos^2\iota)^2\cos^2\theta\sin^2 2\psi\\
        &+\frac{9}{32} \cos^2\iota (1+\cos^2\theta)^2\sin^2 2\psi\\
        &+\frac{9}{8}\cos^2\iota \cos^2\theta \cos^22\psi.
    \end{split}
    \end{equation}
    We take the upper integration bound $f_{\rm max}$ to be $f_{\rm LSO}=\frac{1}{6\sqrt{6}(2\pi)}\frac{c^3}{G(1+z)(m_1+m_2)}$, assumed as the frequency at the last stable orbit. The lower integration bound is $f_{\rm min}=1\rm Hz$.

    Notice that {\it we included the magnification factor $\mu$ in the calculation of the SNR}. This is how we account for selection effects induced by lensing in our analysis.  Fig.~\ref{fig:dndz} shows the redshift and magnification distributions in a realisation of our mock catalog. In the lower panel, we also highlight the difference in the mock catalogue distributions that we obtain by removing $\mu$ from equation~\eqref{SNReq}, and then cutting the events with SNR below the detector threshold.

    In our fiducial scenario, the mock catalogue is comprised of $3,000$ multi-messenger events, which we assume can be collected over a $\sim$10 years operational time of ET, together with EM facilities for the redshift determination \citep{Sathyaprakash2010, Yu2021}. We also consider a more optimistic scenario in which the mock catalogue contains $30,000$ bright sirens.
    
\end{enumerate}

\begin{figure}
    \centering    
    \includegraphics[width=0.45\textwidth]{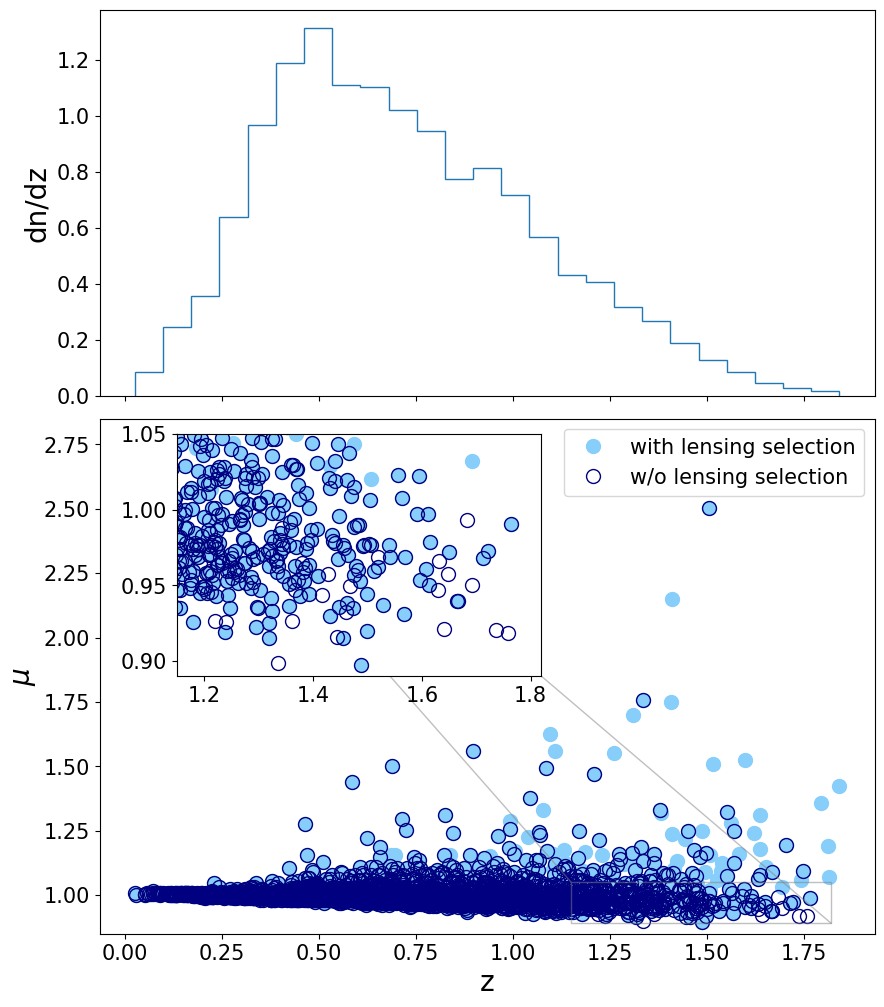}
    \caption{Upper plot: Redshift distribution of the mock catalogue of events. Lower plot: pairs of $\{\hat{z},\hat{\mu}\}$ values. In both plots, we are showing $N_{\rm events}=3,000$, with SNR grater than $\rho_{\rm lim}$. The SNR of the light blue filled circles is calculated including the magnification factor $\mu$, i.e. with equation~\eqref{SNReq}. Conversely, the SNR of the dark blue unfilled circles is independent of the magnification $\mu$. The redshift distributions of the two cases are very similar, so we choose to show the histogram obtained accounting for the lensing selection effect in the upper plot. }
    \label{fig:dndz}
\end{figure}

\subsubsection{$d_{\rm L}$ uncertainty}
We assume that the error on the luminosity distance is given by \citep{Congedo2022},
\begin{equation}
    \sigma_{d_{\rm L}}^2=\sigma_{\rm GW}^2+\sigma_{\rm WL}^2+\bigg(\frac{\partial d_{\rm L}}{\partial z}\bigg)^2(\sigma_z^2+\sigma_{v_{\rm pec}}^2),
    \label{d_l uncertainty}
\end{equation}
where $\sigma_{\rm GW}$ is the instrumental uncertainty of the gravitational wave detector, $\sigma_{z}=0.001(1+z)$ is the spectroscopic redshift uncertainty and $\sigma_{v_{\rm pec}}=\sqrt{\langle v_{\rm pec}^2\rangle}(1+z)/c$ accounts for the source peculiar velocity with $\sqrt{\langle v_{\rm pec}^2\rangle}=500 \,\rm km \;s^{-1}$. The uncertainty due to weak lensing $\sigma_{\rm WL}$ is estimated in \citet{Hirata2010} by the following fitting formula:
\begin{equation}
    \sigma_{\rm WL}=0.066\bigg(\frac{1-(1+z)^{-0.25}}{0.25}\bigg)^{1.8}d_{\rm L}.
\end{equation}
This uncertainty accounts for the (de)magnification of the luminosity distance inferred from a lensed gravitational wave.
Our goal is to check whether this uncertainty fully absorbs any bias of the cosmological parameters. 

In the analysis of actual data, the instrumental uncertainty $\sigma_{\rm GW}$, together with the uncertainties on all the parameters that enter the GW waveform, is obtained by means of a Bayesian analysis of the GW likelihood. A quicker and efficient, although approximate way to assess the measurement accuracy of ET is to use a Fisher matrix analysis of all the GW parameters, e.g. see \citet{Pieroni2022}. Since for our analysis only the uncertainty on the luminosity distance is needed, we make some simplifying assumptions on the value of $\sigma_{\rm GW}$: for our base model we assume that $\sigma_{\rm GW}=0.1 d_{\rm L}$. We also consider a more optimistic value of $\sigma_{\rm GW}=0.02 d_{\rm L}$ \citep{Congedo2022}. Moreover, if we assume that the uncertainties on $d_{\rm L}$ are uncorrelated with the errors of all the other GW parameters, we can use the formula $\sigma_{\rm GW}=d_{\rm L}/\rho$, which can be found inverting the Fisher matrix and marginalising over all the other GW parameters. This scenario is realistic in the cases when the degeneracy between luminosity distance and inclination angle can be broken, for example by the coincident detection of a GRB event. On the contrary, if the degeneracy is present, a more reliable estimate of the uncertainties is given by $\sigma_{\rm GW}=2 d_{\rm L}/\rho$ \citep{Dalal2006, Belgacem2019}.

\subsection{Lensing PDF}
\label{subsec:mu PDF}
Many different simulation suites, fitting formulae and analytical models of the PDF of $\kappa$ and $\mu$ are available in the literature \citep[see][as some examples]{Das2006, Boyle2021,Fosalba2014,Xavier2016,Takahashi2017,Gouin2019,Osato2021}. We are interested in quantifying how modelling choices in the lensing PDF can influence our results. The importance of utilising an accurate magnification PDF in the framework of standard sirens was already highlighted in \citet{Shang2011}. The tail of highly magnified events in particular is envisaged to be difficult to capture. The resolution of the convergence maps, for example, can influence the tail and the skewness of the distribution. In fact, lowering the resolution of convergence maps has the effect of losing the rare overdensities that produce the high magnified events. Similarly, log-normal models have also the tendency of underestimating the probability of high magnifications.
We also compare the results obtained with the N-body simulations with a simple analytical log-normal model.

\subsubsection{Log-Normal analytical model}

We consider a probability distribution for the magnification $\mu$, as given in \citet[Appendix C]{Martinelli2022}. The shape of the distribution at fixed redshift is:
    \begin{equation}
        p(\mu)=\frac{1}{\sqrt{2 \pi}\sigma (\mu-\mu_{\rm min})}\exp\bigg[-\frac{(\ln(\mu-\mu_{\rm min})-m)^2}{2\sigma^2}\bigg].
        \label{log-normal}
    \end{equation}
    The parameters $\sigma$ and $m$ of the distribution are redshift dependent.
    For a magnification distribution with mean $\langle \mu \rangle$ and variance $\sigma^2_{\mu}$, the parameters of the log-normal distribution are:
    \begin{equation}
    \sigma(z)=\sqrt{\log\bigg(\frac{\sigma^2_{\mu}(z)}{(\langle \mu \rangle-\mu_{\rm min}(z))^2}+1\bigg)},
    \end{equation}
    \begin{equation}
        m(z)=\log(\langle \mu \rangle-\mu_{\rm min}(z))-\frac{\sigma^2(z)}{2}.
    \end{equation}
    We assume for the base model $\langle \mu \rangle=1$ and  $\sigma^2_{\mu}(z)=4 \sigma_{\kappa}^2(z)$, where the variance of the convergence is computed from the convergence power spectrum at fixed redshift, obtained from \textsc{pyccl 2.6.1} \footnote{\label{foot:Pyccl}\url{https://github.com/LSSTDESC/CCL}} \citep{Chisari2019}. $\mu_{\rm min}$ is the minimum magnification corresponding to the case when light propagates in the empty space. It is calculated by:
    \begin{equation}
        \mu_{\rm min}(z)=\bigg(\frac{d_{\rm L}(z)}{c(1+z)\lambda(z)}\bigg)^2, \;\;\; \lambda(z)=\int_0^z dz\frac{1}{(1+z)^2H(z)}.
    \end{equation}

\section{Analysis method}
\label{sec:Method}

We perform a parameter estimation analysis, using our mock catalogue of bright sirens, and study the results obtained neglecting the impact of lensing. We then assess whether the fiducial values of the cosmological parameters are correctly recovered, or whether lensing introduces a bias.

\subsection{Fisher matrix}
\label{subsec:Fisher}

Applying a Fisher matrix framework \citep[see][section 3]{Euclidcolab2020}, we combine the information from multiple GW standard sirens to compute the expected uncertainties on the cosmological parameters $\Vec{\theta}=(H_0,\Omega_{\rm m}, ...)$. In principle, the cosmological parameters should be inferred jointly with the astrophysical parameters which describe the BNS population. For simplicity, we keep the latter fixed, and only investigate the effect of lensing on the cosmological parameters. We adopt the following Fisher matrix \citep{Zhao2011}:
\begin{equation}
    \Gamma_{ij}=\sum_{\rm events} \frac{1}{\sigma_{\rm d_{\rm L}}^2(z)}
\frac{\partial d_{\rm L}(z)}{\partial \theta_i}\frac{\partial d_{\rm L}(z)}{\partial \theta_j},
\label{Fisher dl}
\end{equation}
which neglects the dependence of the error $\sigma_{\rm d_{\rm L}}(z)$ on the cosmological parameters. 

Inverting the Fisher matrix $\Gamma_{ij}$, we obtain the errors on the cosmological parameters, $\theta_i$, by,
\begin{equation}
    \sigma_{\theta_i}=\sqrt{\big[\Gamma^{-1}\big]_{ii}}
    \label{marginal errors}
\end{equation}
This error includes the contribution due to the degeneracies with all the other parameters of the Fisher matrix. If we assume that one parameter is perfectly known, we can remove its corresponding row and column from the Fisher matrix, before inverting it. 

\subsubsection{Systematic error}
\label{subsubsec:Amara}
We estimate the bias introduced by neglecting the lensing by employing the formalism of \citet{Amara2008}. In our case, the systematic error comes from the biased luminosity distance inferred from the GW event:
\begin{equation}
    \Delta d_{\rm L}^{\rm sys}=d_{\rm L}^{\rm bias}-d_{\rm L}=\bigg(\frac{1}{\sqrt{\mu}}-1\bigg) d_{\rm L}.
\end{equation}
We compute the ``bias vector'' as:
\begin{equation}
    B_j=\sum_{\rm events}\frac{1}{\sigma^2_{d_{\rm L}}}\Delta d_{\rm L}^{\rm sys} \frac{\partial d_{\rm L}}{\partial \theta_j}.
    \label{bias vector}
\end{equation}
Finally, the bias on the cosmological parameters is obtained from:
\begin{equation}
    b_{\theta_i}=\hat{\theta}_i-\theta_i^{\rm true}=\big[\Gamma^{-1}\big]_{ij}B_{j},
\end{equation}
where we are summing over repeated indices. The derivation 
 of these expressions relies on a first order Taylor expansion of the likelihood, valid for small values of the bias. For larger biases, a study of the full likelihood is more appropriate.

\subsection{MCMC analysis}
For some interesting cases, we check our results with a full Monte Carlo Markov Chain analysis \citep{Hogg2018}. Assuming a Gaussian likelihood for the cosmological parameters $\mathcal{L}(H_0,\Omega_{\rm m})$, we fit our mock catalogue of lensed bright sirens to the theoretical value of the luminosity distance, 
\begin{equation}
    \log \mathcal{L}(H_0,\Omega_{\rm m})=-\frac{1}{2}\sum_{\rm events} \frac{1}{\sigma_{d_{\rm L}}^2}\bigg(d^{\rm bias}_{\rm L}-d_{\rm L}(H_0,\Omega_{\rm m})\bigg)^2,
    \label{MCMCeq}
\end{equation}
where $d^{\rm bias}_{\rm L}$ is constructed with the $\{\hat{z},\hat{\mu}\}$ values in our mock catalog, while the theoretical $d_{\rm L}$ using the redshifts $\{\hat{z}\}$ only. For this, we use the \textsc{emcee} software \footnote{\label{foot:emcee}\url{https://emcee.readthedocs.io/en/stable/}} \citep{Mackey2013}. 

Moreover, to check whether the fiducial values of the cosmological parameters are correctly recovered, we also fit the mock catalogue to the theoretical luminosity distance rescaled by the lensing magnification of each event $\frac{d_{\rm L}(H_0,\Omega_{\rm m})}{\sqrt{\hat{\mu}}}$.

\section{Results}
\label{sec:Results}

We perform the analysis described in Section \ref{sec:Method} using the fiducial realisation of our mock catalog, built as in Section \ref{sec:Mock}. We then consider some variations of the characteristics of the catalog.

\subsection{Fiducial scenario}

\begin{figure*}
    \centering    
    \includegraphics[width=0.75\textwidth]{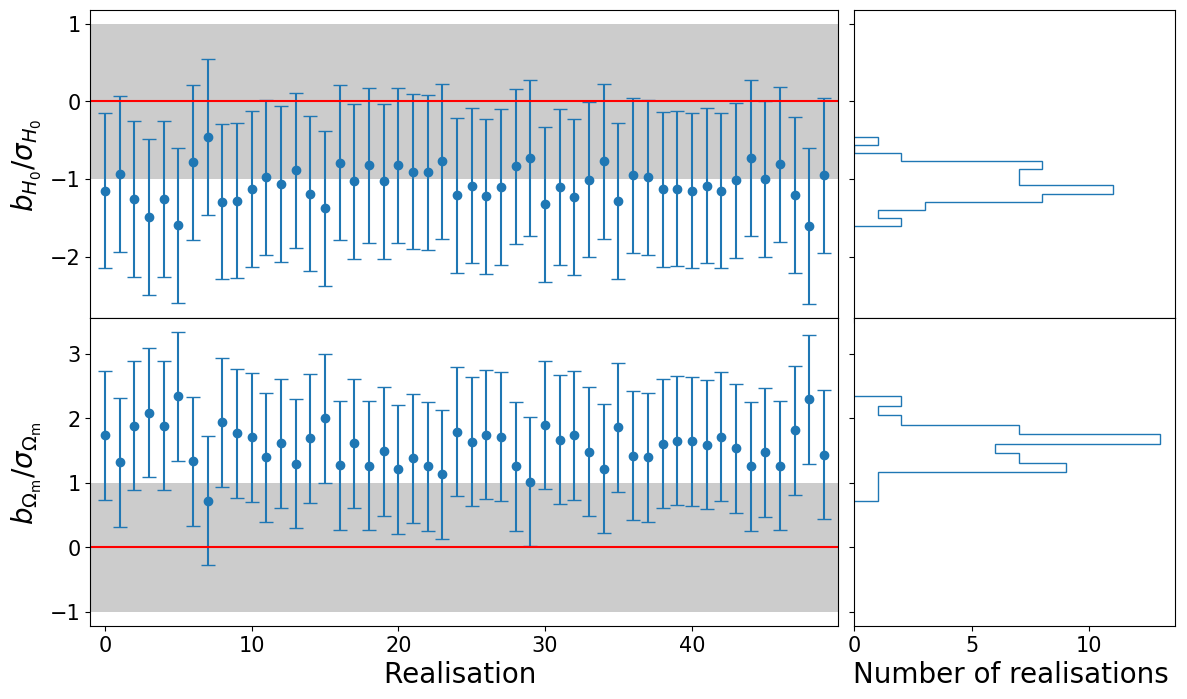}
    \caption{In the left panels: ratio of the bias and statistical error on $H_0$ (upper panel) and $\Omega_{\rm m}$ (lower panel) for 50 different realisations of the mock catalog, each with $N_{\rm events}=3,000$ and assuming $\rho_{\rm lim}=12$, $\sigma_{\rm GW}=0.1 d_{\rm L}$. The gray bands highlight the region of $1\sigma$ significance from zero. The right panels show the histograms of the ratios. }
    \label{fig:BiasH0real}
\end{figure*}

The redshift distribution of the bright sirens events in our fiducial scenario is obtained imposing the EM selection function in equation \eqref{EMsel}, which introduces a maximum redshift of $z=2$, and we assume an instrumental uncertainties on the luminosity distance of $\sigma_{\rm GW}=0.1d_{\rm L}$. In this scenario, we draw the binary inclination angle from an isotropic distribution in the range $[0,\pi]$, thus allowing the possibility to detect off-axis events.  We then impose the detection of beamed GRB events by restricting $ \iota \lesssim 20 \degr$. 

As a first step, we constrain the parameter space of the Hubble constant and the matter density parameter: $(H_0, \Omega_{\rm m})$, which are the only parameters relevant for standard sirens within a flat $\Lambda \rm CDM$ model. We report the results for 50 different realisations of the mock catalog. For each realisation, we draw again the distributions of redshifts, magnifications, angles and masses. The result is obtained using the Fisher matrix analysis presented in Subsection \ref{subsec:Fisher}, with the derivatives in equation \eqref{Fisher dl} and \eqref{bias vector} calculated analytically for stability. While we find that the statistical errors, obtained from the inverse of the Fisher matrix, are stable for different realisations of the mock catalog, the bias fluctuates over more than one order of magnitude. (As more events are considered, the value of the bias becomes more stable, but here we are restricted to $3,000$ in the fiducial catalogue.)

At $1\sigma$ level, the statistical accuracies on $H_0$ and $\Omega_{\rm m}$ are respectively $0.5 \%$ and $3.8\%$. This is comparable or slightly higher than the results obtained in \citet{Cai2017,Belgacem2019, Jin2020}, but we emphasise that they are making different assumptions when constructing the mock catalog, in particular with a different way of calculating the instrumental uncertainty on the luminosity distance and a different number of events, which justify the difference.

In the left panel of Fig.~\ref{fig:BiasH0real}, we show the ratio between the value of the bias and the statistical error on $H_0$ (upper panel) and $\Omega_{\rm m}$ (lower panel), for each realisation of the mock catalog. It is worth noting that the error bars reported in this and the following plots represent the marginalised uncertainties on the corresponding cosmological parameters $\sigma_{\theta_i}$, obtained from the inverse of Fisher matrix, as in equation \eqref{marginal errors}. In the right panels of Fig.~\ref{fig:BiasH0real}, we display the histograms obtained with the ratios of the bias and the statistical uncertainty. We find that lensing induces a negative (positive) systematic error for $H_0$ ($\Omega_{\rm m}$); in fact, for both the cosmological parameters, the values of the bias do not oscillate around zero. While for $H_0$, we find that the absolute mean value of bias has the same order of magnitude than the statistical uncertainty, for $\Omega_{\rm m}$ the peak of the distribution of the ratio $b_{\Omega_{\rm m}}/\sigma_{\Omega_{\rm m}}$ is around $\sim 1.6$, signaling that, in this scenario, the systematic error induced by lensing can dominate the statistical one. 

The amplification of the SNR for magnified events makes them more likely to exceed the SNR threshold with respect to the demagnified ones. This can also be noted in the lower panel of Fig.~\ref{fig:dndz}. The lensing selection effect tends to be more important at high redshift, when the probability for highly magnified events increases. Magnified events look closer than they are, and their biased luminosity distance is better fit by higher values of $\Omega_{\rm m}$. We check the amount of contribution to the bias given by the lensing selection effect. To do so, we set $\mu=1$ in the calculation of the SNR, in equation \eqref{SNReq}, which corresponds to neglecting the selection effect of lensing, and we find that the bias on both $H_0$ and $\Omega_{\rm m}$ is reduced to a value equal or less than $1\sigma$ for all realisations of the mock catalog. Furthermore, removing the lensing selection effect, for the different realisations we find that the bias fluctuates around zero, showing both negative and positive values; as a consequence the average ratios becomes $b_{H_0}/\sigma_{H_0} \sim -0.2$ and $b_{\Omega_{\rm m}}/\sigma_{\Omega_{\rm m}}\sim 0.06$. We conclude that the bias induced by the selection effect alone contributes the most to our results.

\subsubsection{Parameter space}

\begin{figure*}
    \centering    
    \includegraphics[width=0.75\textwidth]{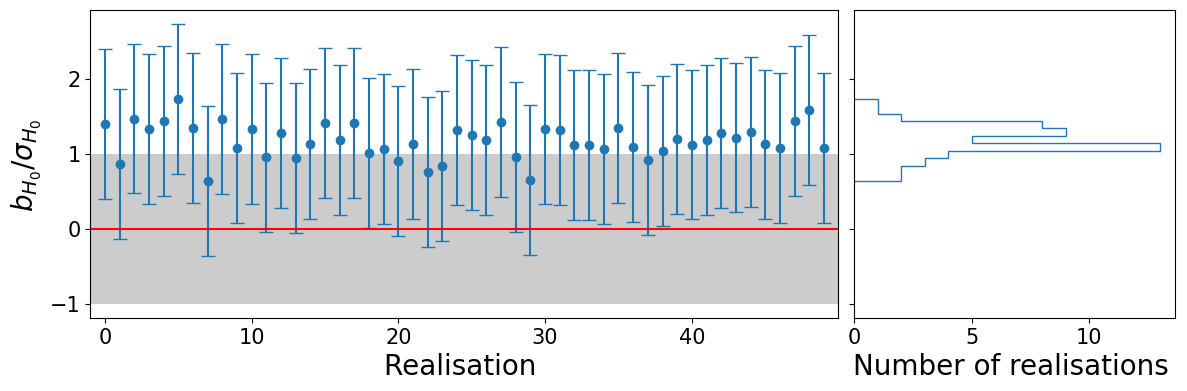}
    \caption{In the left panel: ratio of the bias and statistical error on $H_0$, assuming that $\Omega_{\rm m}$ is perfectly known, for 50 different realisations of the mock catalog, each with $N_{\rm events}=3,000$ and assuming $\rho_{\rm lim}=12$, $\sigma_{\rm GW}=0.1 d_{\rm L}$. The gray bands highlight the region of $1\sigma$ significance from zero. The right panel shows the histogram of the ratios.}
    \label{fig:BiasOmreal_marg}
\end{figure*}

In Fig.~\ref{fig:BiasOmreal_marg}, we show the results obtained with the same set of realisations of the mock catalog, but assuming $\Omega_{\rm m}$ is perfectly known. This is a common assumption in recent studies of standard sirens using the real data of LIGO/Virgo \citep{Abbott2017, Palmese2020}. In our 3G scenario, we can imagine that the matter density parameter will be already very well constrained by other EM probes, while the Hubble constant might still suffer of some residual tension. It is then interesting to check at what level lensing could impact the Hubble tension. As expected, we find that the absolute magnitude of the bias is slightly larger than in the previous case, with a peak of the distribution in the right panel of Fig.~\ref{fig:BiasOmreal_marg} around $\sim 1.2$. Unlike the previous case, the bias on $H_0$ is positive. This is consistent with the fact that, due to the selection effect induced by lensing, the average magnification of the events in the catalogue is slightly greater than one, such that the biased luminosity distance $d_{\rm L}^{\rm bias}$ tends to be smaller than the true one. Given that $\Omega_{\rm m}$ is fixed, the freedom to better fit the data is only left to $H_0$, which is inversely proportional to $d_{\rm L}$, as in equation \eqref{d_l}. This creates a positive bias on $H_0$.

\begin{figure*}
    \centering    \includegraphics[width=0.75\textwidth]{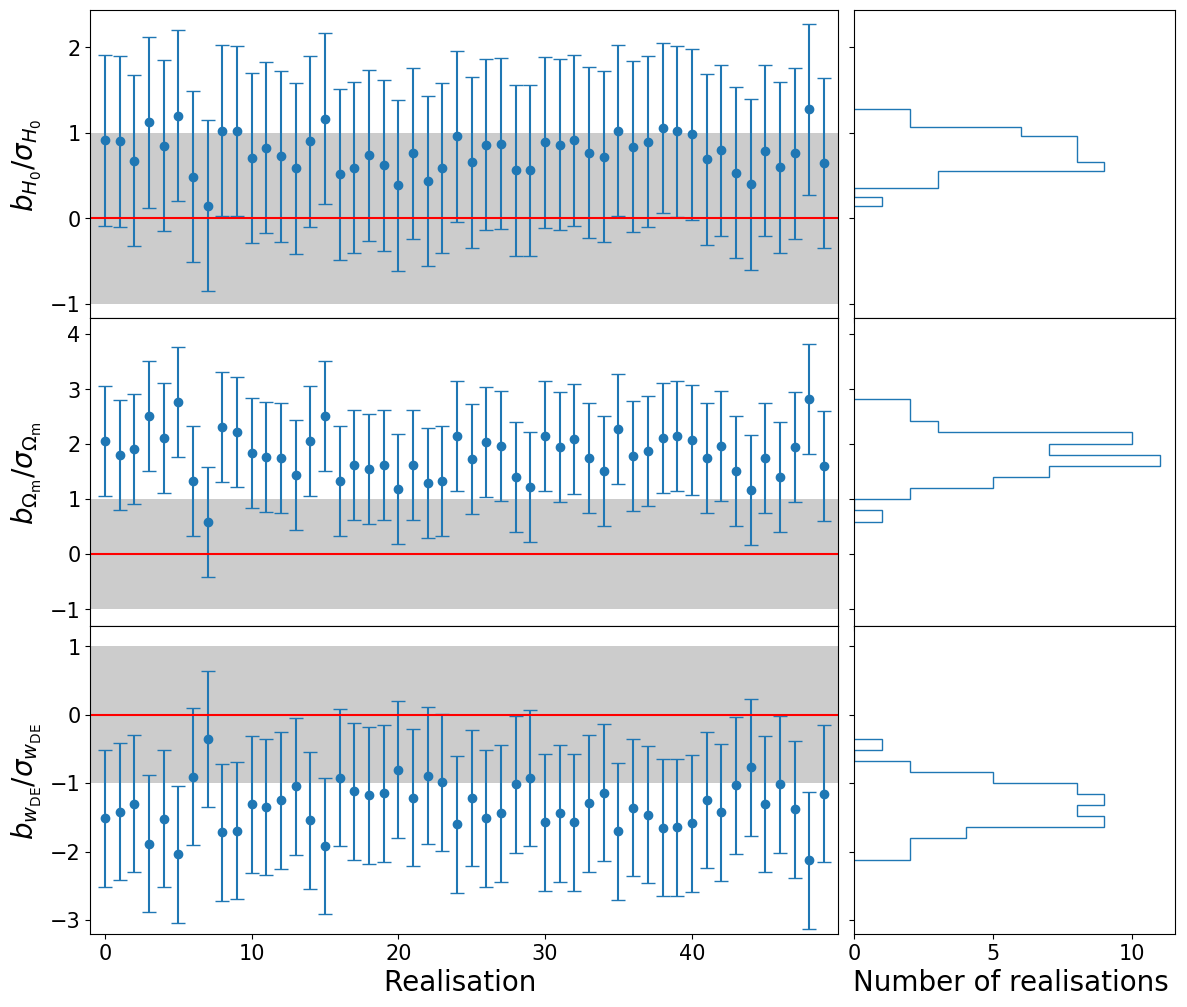}
    \caption{In the left panels: ratio of the bias and statistical error on $H_0$ (upper panel) and $\Omega_{\rm m}$ (central panel) and $w_{\rm DE}$ (lower panel) for 50 different realisations of the mock catalog, each with $N_{\rm events}=3,000$ and assuming $\rho_{\rm lim}=12$, $\sigma_{\rm GW}=0.1 d_{\rm L}$. The gray bands highlight the region of $1\sigma$ significance from zero. The right panels show the histograms of the ratios.}
    \label{fig:BiasH0_Om_wde}
\end{figure*}

Finally, we consider the possibility to use GWs to test extensions of the $\Lambda \rm CDM$ model, in particular the nature of dark energy. Our fiducial value of the dark energy equation of state parameter is $w_{\rm DE}=-1$, which corresponds to the case of a cosmological constant. We are interested in knowing whether lensing can prevent us from recovering the correct value of $w_{\rm DE}$. We underline that our aim is not studying the impact of lensing in alternative theory of gravity, since the lensing simulations suite \citet{Hadzhiyska2023}, we are using, was carried in a $\Lambda \rm CDM$ cosmology. In this work, we simply check whether lensing can bias the fiducial values of $w_{\rm DE}$ within a $\Lambda \rm CDM$ cosmology, thus risking drawing incorrect conclusions from the data. We show the result of this analysis in Fig.~\ref{fig:BiasH0_Om_wde}, for the same 50 realisations of the mock catalog. We find that the average ratios are $b_{H_0}/\sigma_{H_0}\sim 0.8$, $b_{\Omega_{\rm m}}/\sigma_{\Omega_{\rm m}}\sim 1.8$ and $b_{w_{\rm DE}}/\sigma_{w_{\rm DE}}\sim -1.3$.  As can be seen from the lower panel in Fig.~\ref{fig:BiasH0_Om_wde}, for some realisations of the catalog, the lensing induced bias on $w_{\rm DE}$ can have a significance up to $2\sigma$, highlighting the necessity that additional care is needed in order to use GWs to shed light on the nature of dark energy. Once again, we also notice that by changing the parameter space, we find that the bias moves to different directions, being now positive for both $H_0$ and $\Omega_{\rm m}$ and negative for $w_{\rm DE}$. While the bias on $H_0$ and $\Omega_{\rm m}$ tends to decrease the value of the luminosity distance, especially at high redshifts, the bias on $w_{\rm DE}$ moves in the opposite direction due to the correlation with $\Omega_{\rm m}$. 

\subsection{Variations from the fiducial scenario}
\label{subsec:variations}

To assess the impact of our assumptions on the previous results, we vary one assumption at a time and leave all the other characteristics of the mock catalogue equal to what was described for the fiducial scenario. 

\begin{figure}
    \centering    
    \includegraphics[width=0.4\textwidth]{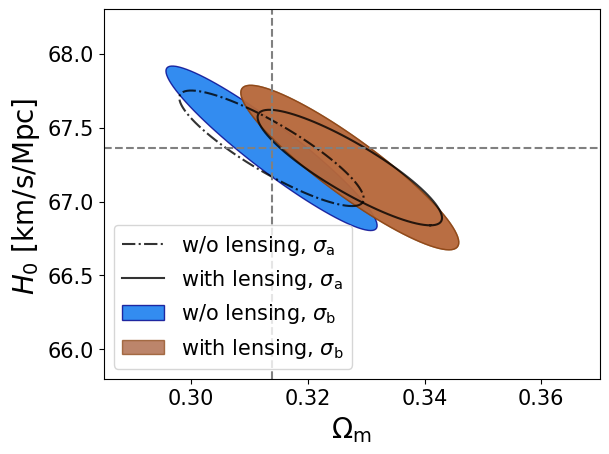}
    \caption{$1\sigma$ ellipses for $H_0$ and $\Omega_{\rm m}$, calculated using the realisation $r_0$ of the mock catalogue and changing the assumptions on the instrumental uncertainty on $d_{\rm L}$: $\sigma_{\rm a}=2d_{\rm L}/\rho$ and $\sigma_{\rm b}=0.1 d_{\rm L}$. The fiducial ellipses, labelled with "w/o lensing", are centered at the correct cosmology, while the biased ones ("with lensing") show the systematic displacement. The intersection of the dotted lines represents the fiducial values of the parameters in standard Cosmology. }
    \label{fig:dl_uncertainty}
\end{figure}

We start from the instrumental uncertainties on $d_{\rm L}$, and we repeat the analysis assuming that $\sigma_{\rm GW}=2d_{\rm L}/\rho$, as done in \citet{Dalal2006}. In this analysis, we pick one realisation of the mock catalog, chosen between the 50 different realisations already studied in the fiducial scenario, as a representative example; we consider in particular the realisation corresponding to the first point in the left panels of Fig.~\ref{fig:BiasH0real}, which for the remainder of the discussion we call $r_0$. Assuming $\sigma_{\rm GW}=0.1 d_{\rm L}$, we obtained $b_{H_0}/\sigma_{H_0}\sim-1.2$ and $b_{\Omega_{\rm m}}/\sigma_{\Omega_{\rm m}}\sim1.7$. Instead, with the  choice $\sigma_{\rm GW}=2d_{\rm L}/\rho$ we find that the statistical error slightly decreases to $0.4\%$ and $3.2\%$ respectively for $H_0$ and $\Omega_{\rm m}$, and similarly happens for the absolute values of the systematic bias. As a consequence, for this specific realisation of the mock catalogue with $\sigma_{\rm GW}=2d_{\rm L}/\rho$ the ratios decrease to $b_{H_0}/\sigma_{H_0}\sim-0.5$ and $b_{\Omega_{\rm m}}/\sigma_{\Omega_{\rm m}}\sim1.3$. In Fig.~\ref{fig:dl_uncertainty}, we display the comparison between the $1\sigma$ Fisher confidence ellipses obtained utilising the two different assumptions on $\sigma_{GW}$ discussed above. Notice that the fiducial ellipses are centered at the point corresponding to our fiducial cosmology, while the biased ellipses are shifted by the lensing bias, calculated as in Subsection~\ref{subsubsec:Amara}. Alternatively, one might consider more optimistic scenarios for the instrumental uncertainty on the luminosity distance, for example assuming $\sigma_{\rm GW}=d_{\rm L}/\rho$, which corresponds to the Fisher matrix error obtained marginalising over all the other GW parameters. In this case, we find $b_{H_0}/\sigma_{H_0}\sim-1.1$ and $b_{\Omega_{\rm m}}/\sigma_{\Omega_{\rm m}}\sim2.3$. Finally, adopting $\sigma_{\rm GW}=0.02 d_{\rm L}$ \citep{Congedo2022}, we obtain the highest values of $b_{H_0}/\sigma_{H_0}\sim-2.5$ and $b_{\Omega_{\rm m}}/\sigma_{\Omega_{\rm m}}\sim3.8$.

\begin{table}
	\centering
	\caption{Values of the average ratios of $b_{H_0}/\sigma_{H_0}$, $b_{\Omega_{\rm m}}/\sigma_{\Omega_{\rm m}}$ and $1\sigma$ uncertainties, for three variations of our fiducial scenario: removing the EM selection function given in eq.~\eqref{EMsel} (first row), assuming a uniform mass distribution for the two NSs the the binary system (second row) and restricting the inclination angle to be smaller than $20 \degr$ (third row). The results are obtained averaging over 50 realisations of the mock catalog.}
	\label{tab:Variation}
	\begin{tabular}{lcccr} % four columns, alignment for each
		\hline
		 Model & $\frac{b_{H_0}}{\sigma_{H_0}}$ & $\sigma_{H_0}$ [km/s/Mpc] & $ \frac{b_{\Omega_{\rm m}}}{\sigma_{\Omega_{\rm m}}}$ & $\sigma_{\Omega_{\rm m}}$ \\
		\hline
		w/o $f(z)$ & -1.5 & 0.4 & 2.1 & 0.01 \\
		NS $p(m_{1/2})$ uniform & -0.7 & 0.4 & 1.1 & 0.01\\
		$\iota<20 \degr$ & -1 & 0.4 & 1.4 & 0.01 \\
		\hline
	\end{tabular}
\end{table}

We also explore how some assumptions regarding the probability distributions of parameters that characterise the GW event impact our results. We draw 50 new realisations of the mock catalogue and quote the average bias in Table~\ref{tab:Variation}. Future high-energy detectors, such as THESEUS \citep{Amati2018,Stratta2018} and TAP \citep{Camp2017}, are expected to detect GRBs at redshifts higher than $\sim 2$. Moreover, given the expected timeline of 3G detectors, another beyond-Euclid generation, with higher-redshift reach, of spectroscopic surveys may start to be available \citep{Schlegel2022}. Therefore we might assume that all the events in our simulated catalogue will have, besides the GRB emission, a detectable afterglow, useful for the redshift determination. This would justify removing the EM selection function in equation \eqref{EMsel}. The resulting redshift distribution shows a tail of events up to redshift $\sim 3$. As given in Table~\ref{tab:Variation}, on average, the absolute values of the bias increases for both cosmological parameters, $H_0$ and $\Omega_{\rm m}$. This is expected since the few events in the tail of high redshift do not properly sample the lensing PDF and, in addition, they are more likely to be highly magnified. Given that the AbacusSummit simulations cover up to $z=2.45$, we have to extrapolate the PDF to higher redshifts. Since we have only a few events above this maximum redshift of $z=2.45$, they should not be affected by how we choose to perform the extrapolation. Moreover, we consider a more pessimistic scenario in which the EM observations favor only lower-distance observations, i.e. we decrease the redshift cut-off in equation \eqref{EMsel} to  $z_{\rm pivot}=1.3$ and $z_{\rm max}=1.5$, such that no events can be detected above this $z_{\rm max}$. We find that the significance of the bias becomes smaller than $1\sigma$ for both the parameters, in particular we obtain $b_{H_0}/\sigma_{H_0}=-0.5$ and $b_{\Omega_{\rm m}}/\sigma_{\Omega_{\rm m}}=0.8$. This is expected since the impact of lensing is less significant at smaller redshifts, as can be seen from the magnification PDF in Figure~\ref{fig:pmu}. 

We check the dependence of our results on the mass distribution of the NS in the binary system. The fiducial scenario assumed a Gaussian distribution for both the components of the binary, as explained in Sec.~\ref{sec:Mock}. We draw the value of the two masses from a uniform distribution between $[1;2.5]\rm M_{\sun}$ and display in Table~\ref{tab:Variation} the average bias. The result is similar to the case of a Gaussian distribution for the NS masses, with a slight average decrease in the bias both on $H_{0}$ and $\Omega_{\rm m}$ of a factor $0.7$.

Furthermore, we restrict the value of the inclination angle $ \iota \lesssim 20 \degr$, assuming that only beamed GRBs can be detected. This case is reported in the last row of Table~\ref{tab:Variation}. The average lensing bias on both the cosmological parameters remains stable. On the other hand, considering a more optimistic scenario with $\sigma_{\rm GW}=0.02 d_{\rm L}$ and $\rho_{\rm \lim}=8$ (more details on this scenario are given in Appendix~\ref{AppendixA}), the average lensing bias on both the cosmological parameters tends to decrease, in particular of a factor $0.6$ for $H_0$ and $0.5$ for $\Omega_{\rm m}$. This is expected, since on-axis events are louder, and thus less affected by lensing selection effects. We note that the choice $ \iota < 20 \degr$ might be conservative. For example, the detection of the radio afterglow may allow to relax the previous constraint \citep{Dobie2021}. We consider the two scenarios, $\iota$ drawn from an isotropic distribution and $\iota<20 \degr$, as two extreme cases.

\subsubsection{Log-Normal analytical model}

\begin{figure}
    \centering    
    \includegraphics[width=0.4\textwidth]{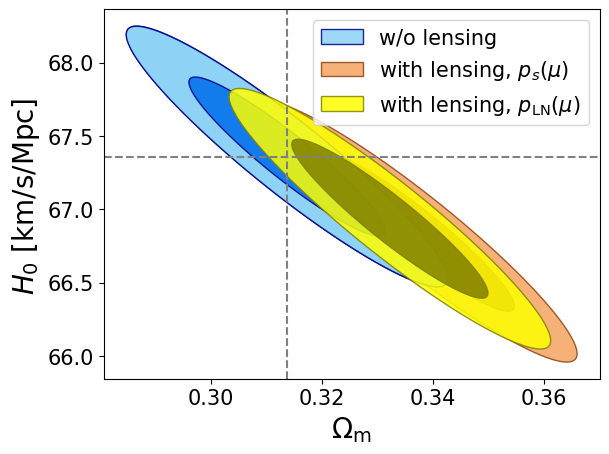}
    \caption{$1\sigma$ and $2\sigma$ ellipses for $H_0$ and $\Omega_{\rm m}$, calculated drawing the lensing magnifications from the log-normal model $p_{\rm LN}(\mu)$ (yellow ellipse) and the simulations $p_{\rm s}(\mu)$ (orange ellipse). The fiducial ellipse, labelled with "w/o lensing", are centered at the correct cosmology, while the biased ones ("with lensing") show the systematic displacement. The intersection of the dotted lines represents the fiducial values of the parameters in standard Cosmology.}
    \label{fig:LNvsSim}
\end{figure}

We repeat the analysis using the analytical log-normal model, presented in Sec.~\ref{subsec:mu PDF}. The shape of this magnification PDF, with respect to the one derived from simulations, differs the most around the peak of the distribution, where most of the events fall. In fact, the shape of PDF from simulations is more sharply peaked around $\mu\sim 1$, such that demagnified events have higher values of the magnification. The tail of the two distributions also shows differences, in particular, being underestimated in the log-normal model, where the probability of highly magnified events ($\mu \gtrsim 1.5$) is smaller. We compare the results obtained from the same realisation, $r_0$, of the mock catalogue but with different magnification values. We use the same distributions for redshifts, masses and angles, while we draw the magnifications from the two different PDF, sampled by the same random numbers. We then calculate the SNR \eqref{SNReq} of the events in the two cases and keep only the ones that exceed the threshold. In Fig.~\ref{fig:LNvsSim}, we show the confidence ellipses obtained with this analysis.

For our fiducial scenario, we find that the values of systematic bias of $H_0$ and $\Omega_{\rm m}$ calculated using the log-normal model are very similar to the ones obtained with the simulations. On the other hand for the more optimistic scenario in which we consider $\sigma_{\rm GW}=0.02d_{\rm L}$ and $\rho_{\rm lim}=8$, the values of systematic bias of $H_0$ and $\Omega_{\rm m}$ calculated using the log-normal model are less than half the ones obtained with the simulations, in particular with a decrease of a factor $\sim 0.4$ for $H_0$ and $\sim 0.3$ for $\Omega_{\rm m}$. A similar decrease in the cosmological parameters bias can be found by utilizing the PDF from simulations and removing the events with $\mu \gtrsim 1.4$. This result highlights the importance of an accurate modelling of the magnification PDF, and in particular of its tail, which is underestimated in the log-normal model. Note that other works have found log-normal descriptions of the lensing PDF that are in better agreement with simulations \citep{Marra2013}.
 
Even though we consider the PDF obtained from the N-body simulations to be more reliable, recent simulations do not represent the final truth about the magnification PDF and its tail. Resolution can impact the details of the tail. Moreover, baryonic effects, which are not included in the simulations we use, may also have an impact. According to \citet[Figure 5]{Osato2021}, baryons affect the convergence field distribution at $\sim 10\%$ level, with a smaller impact at higher redshifts.

\subsubsection{Number of events}

\begin{figure}
    \centering    
    \includegraphics[width=0.4\textwidth]{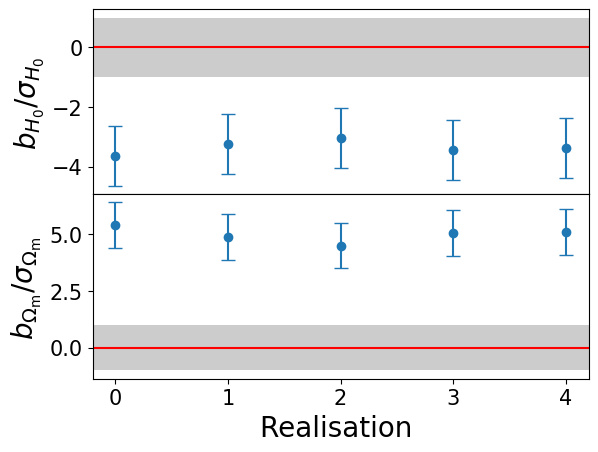}
    \includegraphics[width=0.4\textwidth]{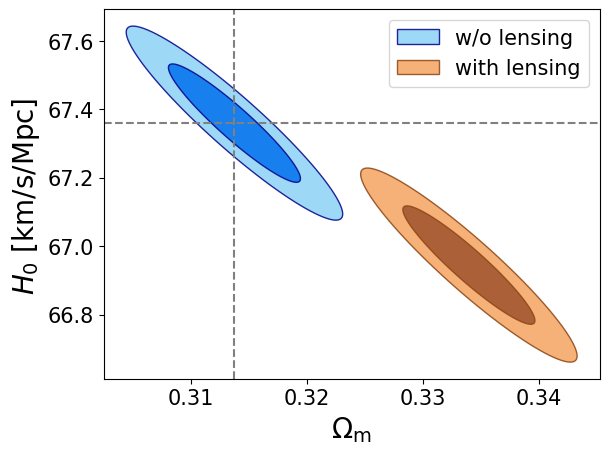}
    \caption{Upper plot: ratio of the bias and statistical error on $H_0$ and $\Omega_{\rm m}$ for 5 different realisations of the mock catalog, $N_{\rm events}=30,000$. The gray bands highlight the region of $1\sigma$ significance from zero. Lower plot: ellipses showing the $1\sigma$ and $2\sigma$ contours obtained considering the realisation $r_0'$ of the mock catalog.}
    \label{fig:Bias_30000}
\end{figure}

Finally, we discuss an optimistic scenario in which we consider a factor of 10 more events, i.e. we build a mock catalogue with $N_{\rm events}=30,000$. The statistical error, at $1\sigma$ level, reduces to $0.1 \%$ for $H_0$ and to $1.3 \%$ to $\Omega_{\rm m}$, slightly tighter than the current 
EM constraints, for example by \citet{Planck2018}. This high level of accuracy is achieved keeping the same assumption as in the previous case on the instrumental errors on $d_{\rm L}$, i.e. $\sigma_{\rm GW}=0.1d_{\rm L}$. Even though these numbers are probably too optimistic even for 3G detectors, we find this idealised scenario suitable to show the effect of the systematic bias. In fact, for different realisations of the mock catalogue the dispersion of the bias around the mean turns out to be smaller, as it is clear from the upper panel of Fig. \ref{fig:Bias_30000}, although a smaller number of realisations is considered. Moreover, under these assumptions, the systematic bias is not compatible with zero, with a significance of more than $3\sigma$ for $H_0$ and of about $5\sigma$ for $\Omega_{\rm m}$. In the lower panel of Fig.~\ref{fig:Bias_30000}, the confidence $1\sigma$ and $2\sigma$ ellipses for one realisation of the mock catalogue are shown. Similarly as before, we choose the first point of the upper panel of Fig.~\ref{fig:Bias_30000} as a representative example and we called it $r_0'$. Also in this scenario, we find that the lensed events are better fitted by smaller (higher) values of $H_0$ ($\Omega_{\rm m}$).

On the contrary, with the observation of a smaller number of bright siren events, the significance of the bias decreases. We consider the case of $N_{\rm events}=300$, so 30 multi-messenger events per year. In this case, we find that $b_{H_0}/\sigma_{H_0}=-0.4$ and $b_{\Omega_{\rm m}}/\sigma_{\Omega_{\rm m}}=0.6$. In fact, with less events the statistical uncertainty on the cosmological parameters increases and absorbs the bias.

\subsection{MCMC validation}

\begin{figure}
    \centering    
    \includegraphics[width=0.4\textwidth]{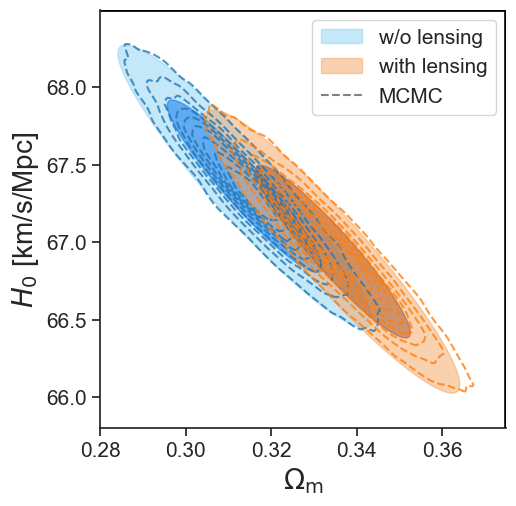}
    \caption{Confidence contours obtained with the Fisher formalism ($1\sigma$ and $2\sigma$ filled ellipses) and the MCMC analysis (dotted lines). The contours are obtained using the same base realisation of the mock catalogue with $N_{\rm events}=3,000$.}
    \label{fig:MCMCvsFisher}
\end{figure}

All the results showcased in previous sections were performed with the Fisher matrix formalism. However, as explained in Section \ref{sec:Method}, we validate the results with the full likelihood with a MCMC analysis. We compare the outcome of the two methods for a few representative cases. In Fig. \ref{fig:MCMCvsFisher}, we display the comparison for the fiducial scenario, corresponding to the realisation $r_0$ of the mock catalog. As can be seen from the plot, the MCMC contour lines match the Fisher ellipses with very good agreement. Given that we expect the Fisher approach to fail for high value of the bias, we test the result for a realisation of the mock catalogue showing one of the biggest systematic bias among those observed in our analysis. We consider a realisation of the mock catalog, which we called $r_1$, obtained removing the EM selection function while keeping all the other assumptions unchanged with respect to our fiducial scenario. With the Fisher analysis, we obtain $b_{H_0}=-0.72 \, \rm km/s/Mpc$ with $\sigma_{H_0}=0.35 \, \rm km/s/Mpc$, and $b_{\Omega_{\rm m}}=0.03$ with $\sigma_{\Omega_{\rm m}}=0.01$. As for the previous case, we find perfect agreement with the MCMC result, obtaining the same significant figures. 

Finally, we check our results adding statistical noise in value of the luminosity distances in our mock catalog. In the previous analysis we fitted the biased luminosity distance to its theoretical value, as given in equation~\eqref{MCMCeq}. However, the measured luminosity distances from actual observations are noisy, in addition to biased. To account for this, we sum to $d_{\rm L}^{\rm bias}$ in equation~\eqref{MCMCeq} a statistical error $\Delta d_{\rm L}$, which we draw from a Gaussian distribution with width equal to the expected uncertainty on $d_{\rm L}$. We repeat the previous analysis, considering again the same realisation $r_0$ of the mock catalog, and drawing different realisations of the noise. We notice that even in the fiducial case, i.e. fitting the mock data to the theoretical value of the luminosity distance re-scaled by the magnification factor, the confidence contours are not perfectly centered at the point corresponding to the fiducial values of the cosmological parameters. In particular, we see that, for different realisations of the noise, the position of the contours are randomly scattered around the point corresponding to the fiducial cosmology, even though always remaining consistent at $1\sigma$ level.  On the other hand, for different realisations of the noise, we find that the value of the bias, computed as the displacement between the mean values of the confidence contours, remains stable and equal to the one obtained without including the statistical noise in the data. 

\section{Mitigation}
\label{sec:Mitigation}

\begin{figure*}
    \centering 
    \includegraphics[width=0.4\textwidth]{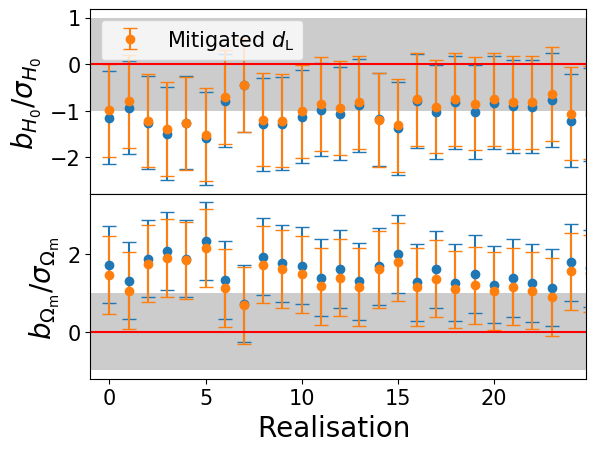}
    \includegraphics[width=0.4\textwidth]{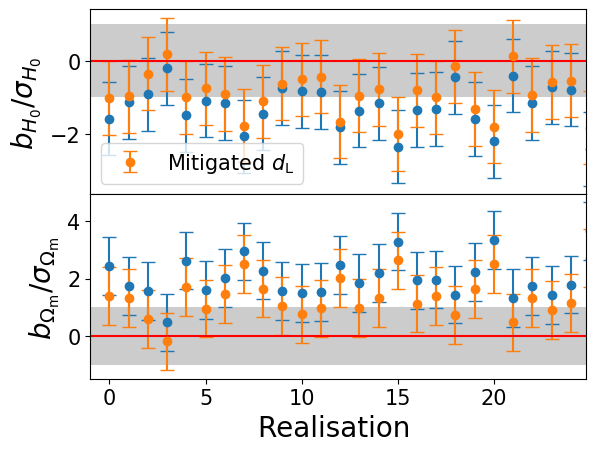}
    \caption{Ratio of the bias and the statistical uncertainty on $H_0$ (upper panel) and $\Omega_{\rm m}$ (lower panel) for 25 different realisations of the mock catalog, assuming that $\sigma_{\rm GW}=0.1 d_{\rm L}$ and $\rho_{\rm lim}=12$ (on the left) and $\sigma_{\rm GW}=0.02 d_{\rm L}$ and $\rho_{\rm lim}=8$ (on the right), $N_{\rm events}=3,000$, with mitigation. The blue points show the absolute value of the bias, without any correction; the orange points show the absolute value of the bias obtained after applying the mitigation on the luminosity distance, as in equation \eqref{mitigation}. The gray bands highlight the region of $1\sigma$ significance from zero.}
    \label{fig:BiasH0real_mitig}
\end{figure*}

    \begin{figure}
    \centering    \includegraphics[width=0.4\textwidth]{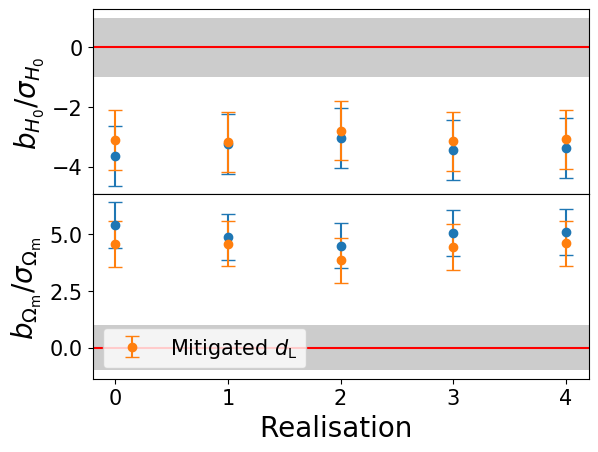}
    \includegraphics[width=0.4\textwidth]{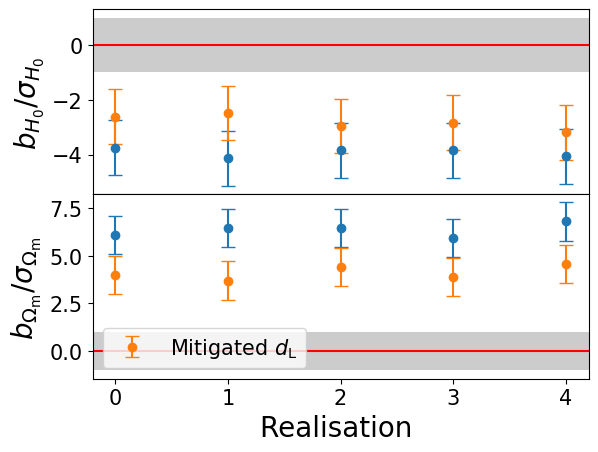}
    \caption{Ratio of the bias and statistical uncertainty on $H_0$ and $\Omega_{\rm m}$ for 5 different realisations of the mock catalog, assuming that $\sigma_{\rm GW}=0.1 d_{\rm L}$ and $\rho_{\rm lim}=12$ (on the top) and $\sigma_{\rm GW}=0.02 d_{\rm L}$ and $\rho_{\rm lim}=8$ (on the bottom). In blue the systematic bias and in orange the one obtained from the mitigated $d_{\rm L}$, $N_{\rm events}=30,000$.}
    \label{fig:Bias5real}
\end{figure}

We would like to find a procedure to mitigate the impact of the lensing bias in the estimation of the cosmological parameters. Previous ``delensing'' attempts in the literature focus on the reduction of statistical errors induced by lensing. The correction is achieved thanks to an independent estimation, which relies on EM observations, of the amount of magnification expected along the line of sight, with an event-by-event procedure \citep{Shapiro2010, Hilbert2011, Wu2023}. Even though a similar strategy could be successfully used to address the issue of systematic errors, we consider an alternative, less costly way to statistically mitigate the lensing impact of a catalogue of bright sirens.

For each event in our mock catalogue we draw from the magnification PDF a second independent value of the magnification, which we call $\mu_{\rm M}$. This procedure requires knowledge of the lensing magnification PDF and of the redshift of each event in the catalogue. In the former case, we can for example rely on the results of ray-tracing simulations, as done in this work, while in the latter case the redshift is known from the EM counterpart of each event. We use $\mu_{\rm M}$ to correct the biased luminosity distance, which is obtained from the parameter estimation of the GW. This correction is applied after the selection of events with SNR greater than the SNR threshold. For each event, the mitigated luminosity distance $d_{\rm L}^{\rm MIT}$ is given by:
    \begin{equation}
        d_{\rm L}^{\rm MIT}=d^{\rm bias}_{\rm L}(\mu) \sqrt{\mu_{\rm M}}=d_{\rm L}^{\rm true}\sqrt{\frac{\mu_{\rm M}}{\mu}},
        \label{mitigation}
    \end{equation}
where $d_{\rm L}^{\rm true}$ is the true luminosity distance of the event. We calculate the bias with the Fisher formalism, using $d_{\rm L}^{\rm MIT}$ instead of the biased luminosity distance. To help reducing fluctuations due to events in the tail of the magnification distribution, we repeat the procedure 10 times and take the average of the mitigated values of the bias for each realisation. The result is shown in Fig. \ref{fig:BiasH0real_mitig}, in the left panel for our fiducial mock catalog, while in the right panel we consider a more optimistic scenario with $\rho_{\rm lim}=8$ and $\sigma_{\rm GW}=0.02 d_{\rm L}$. As can be seen, for almost all the realisations we are able to slightly reduce the systematic error, with different amount of correction. In fact, the mitigation procedure reduces the systematic bias on both the cosmological parameters or in the worst case, it keeps it unchanged. In the fiducial scenario (left panel of Fig. \ref{fig:BiasH0real_mitig}) we find that $b_{H_0}/\sigma_{H_0}\sim -0.96$ and $b_{\Omega_{\rm m}}/\sigma_{\Omega_{\rm m}}\sim 1.37$. This corresponds to an average reduction of a factor of $0.9$ for both $H_0$ and $\Omega_{\rm m}$. An higher level of correction is instead achieved in the optimistic scenario (right panel of Fig. \ref{fig:BiasH0real_mitig}), in which we find an average reduction of a factor of $0.7$ for $H_0$ and $0.6$ for $\Omega_m$. To achieve a complete correction, we think that more accurate methods are needed. As we said, a possibility might be to assume that the lensing magnification of each event is known with a certain precision, thanks to EM observations.

We also show the mitigation in the case of $N_{\rm events}=30,000$, in Fig. \ref{fig:Bias5real}, obtained from single set of $\{\hat{\mu}_{\rm M}\}$. Indeed, given the larger sample, we do not average over multiple values of the mitigated bias. In this case, even if we still observe a reduction of the systematic bias for both the cosmological parameters, in any realisation we are unable to find a mitigation that makes the bias compatible with zero or with a significance equal or less than $1\sigma$. Indeed $b_{H_0}$ and $b_{\Omega_{\rm m}}$ remains incompatible with zero at a significance of more than $3\sigma$ for both scenarios. This is also due to the fact that in this case we have very small error bars on the cosmological parameters, thus more accurate methods are needed for correcting the biased induced by lensing.

\subsection{Additional considerations}
\label{subsec: add}
We also consider alternative approaches that, alone or in combination, can help with the mitigation of the bias (figures are not included). 

We try to remove from our catalogue the highly magnified events from the tail of the distribution. We suppose that events that fall into the strong lensing regime can be recognised and thus not used for the cosmological parameters' inference. Data analysis tools have been developed and they allow to carry strong lensing searches in GW data. These searches aim at successfully pairing different images of the same event \citep{Haris2018, Lo2023} or looking for lensing induced waveform distortions of the single image \citep{Dai2020,Wang2021, Janquart2021, Ezquiaga2021}. We take as reference value $\mu=2$ to separate the weak from the strong lensing regime \citep[see][Figure 15]{Takahashi2011}. For our fiducial scenario, we find that in order to meaningfully remove the bias, i.e. average bias is below $1\sigma$, it is necessary to remove the high magnification tail for $\mu \gtrsim 1.4$. Thus, we believe that the identification of the strongly lensed signals will not suffice to correct for the systematic bias. 

As another attempt, we suppose to know the lensing magnification of the events in the catalogue with a certain precision. For each event, we draw the second value of the lensing magnification $\mu_M'$ from a Gaussian distribution centered around the true magnification $\mu_{\rm true}$. We vary the width of the Gaussian, $\sigma_{\mu_{\rm true}}$, in order to asses the precision required to reduce the bias. We compare the results for one realisation of the mock catalog. Assuming a $10\%$ precision, i.e. $\sigma_{\mu_{\rm true}}=0.1 \mu_{\rm true}$, we find that the bias on $H_0$ increases of more than one order of magnitude, while it remains stable for $\Omega_{\rm m}$. Only assuming that $\mu_{\rm true}$ is known with $\sim 1\%$ precision we can actually reduce it. We found that, the larger the width of the Gaussian, the more the distribution of $\mu_M'$ deviates from our fiducial magnification PDF. Once again this underlines the importance of a correct modelling of the magnification PDF.

Finally, we consider the impact of the choice of the SNR threshold. Increasing $\rho_{\rm lim}$ to $14$, while keeping all the assumptions of the fiducial scenario unchanged, we notice a slight worsening of the bias, in particular for $H_0$ with an increase of a factor $1.2$. This is expected since, with increasing $\rho_{\rm lim}$, demagnified, high redshift events are the first to tend to fall below the threshold, leaving a high redshift tail of magnified events. On the other hand, when the threshold is high enough the mock catalogue only consists of small redshift events, for which the magnification PDF is sharply peaked around $\mu=1$. Clearly, in this limit no lensing bias is expected. For our mock catalog, we find that this condition is met if we impose $\rho_{\rm lim}=25$. In this case, the maximum redshift in the mock catalogue reduces to $z \sim 0.8$ with only a few magnified events at higher redshift, and the bias becomes negligible. On the other hand, only a small number of events survives the SNR cut and so the errors on the cosmological parameters are bigger. Although future investigations are needed, we can suppose that the analysis of the high-SNR subgroup can serve as a sanity check for the results obtained by the full catalogue of detection. We also point out that, with increasing $\rho_{\rm lim}$, much more statistics is needed, since a smaller fraction of events survive the SNR cut. Due to the long computational time, a smaller number of realisations of the mock catalog is considered when increasing $\rho_{\rm lim}$. We expect that the details of this analysis might change with a more complete study, but not the overall conclusions.

\section{Discussion}
\label{sec:Discussion}

We showed that gravitational lensing can induce a systematic bias in the estimation of the cosmological parameters from distances inferred through GWs. We reported the averages over multiple realisations of the mock catalog, since we found that for $N_{\rm events}=3,000$, the value of the systematic bias fluctuates over more than one order of magnitude. Even though in some scenarios, e.g. when restricting  $ \iota \lesssim 20 \degr$, we found that the average bias has a significance smaller than $1\sigma$, it is important to highlight that a few realisations of the mock catalogue still show big fluctuations, up to $2\sigma$ level. It is then of crucial importance to be aware of this in the analysis of real data.

As anticipated, we expect this bias to be sourced by two distinct effects: selection effects and the magnified (or demagnified) luminosity distance. We found that the former effects is dominant for a population of NS binaries, detected by a third-generation GW detector. Previous studies on lensing selection effects were presented in \citet{Cusin2019,CusinTamanini2021, Shan2021}. They define a detector-dependent magnification PDF, that accounts for the finite sensitivity of the detector, and estimate the average magnification expected at a given redshifts. This does not give the probability in absolute terms of observing an event from the given redshift. To do so, we found it more convenient to directly include the magnification in the calculation of the SNR. The SNR cut keeps only the events above the detector's threshold and thus tends to select magnified events whose SNR is increased by a factor $\sqrt{\mu}$. \citeauthor{CusinTamanini2021} present an unbiased estimator for the luminosity distance, which accounts for the average magnification induced by lensing selection effects. We leave for future works to check whether this estimator can mitigate the bias obtained in our analysis. In \citeauthor{Shan2021}, a forecast for ET is reported with the conclusion that no lensing selection effects are expected. The reason is that almost all GWs from the black holes population that they consider can be detected, even without the magnifying effect of lensing. This is not the case in our study, where we consider the NS population accompanied by an EM counterpart; in fact, with increasing redshift, many more events are found close to the SNR threshold.

Another point of comparison can be made with lensing of Type Ia Supernovae (SNe). Indeed, similarly to GWs, calibrated SNe are used as standard candles to test the distance-redshift relation. In \citet{HolzLinder2005}, the lensing degradation of the power of SNe in the estimation of cosmological parameters is investigated. They highlight that lensing can induce a systematic bias, as a consequence of the non-Gaussian shape of the magnification PDF. If it is not well sampled, the mode of the PDF can show an offset from the mean, taken to be unity. In our analysis, this corresponds to the case in which we do not include the magnification in the calculation of the SNR. Similarly to \citet{HolzLinder2005}, we find that lensing bias has negligible impact on the estimation of the cosmological parameters. Moreover, SNe surveys also suffer of selection effects which can spoil the cosmological parameter estimation \citep{Kessler2019}. For example, the Malmquist bias is due to the magnitude limitation of real surveys, which, near the threshold, tends to select brighter events. Lensing selection effects, analogous to the one considered in this work, are also expected for SNe survey, as highlighted in \citet{Shah2023}. They investigate the Pantheon compilation \citep{Scolnic2018}, looking for bias on the cosmological parameters by de-lensing the SNe magnitudes. They find moderate evidence that magnified SNe events were preferentially selected near the magnitude threshold.

We stress that in this work we always assume that the redshift of the GW source is known, thanks to the detection of an EM counterpart. Pairing the GW with the incorrect EM transient will spoil the redshift determination of some of the events. In the absence of an EM counterpart, lensing can also impact the inference of the source-frame masses of the GW sources \citep{He2022}. Moreover, different methods have been proposed to estimate the redshift also in the absence of an EM counterpart. ``Dark sirens'' in fact constitute the great majority of GW events observed so far \citep{Palmese2020, Abbott2021a, Finke2021,Mukherjee2022,Palmese2023, Abbott2023}. In the case of BNS and BHNS systems, the measurement of tidal effects promises to provide an accurate way to break the mass-redshift degeneracy \citep{Messenger2011,Ghosh2022,Shiralilou2022}. Alternatively, the ``galaxy catalogue method''  statistically infers the host galaxy \citep{DelPozzo2012,Fishbach2019,Gray2020}, and the ``spectral siren method'' exploits the relation between the source-frame and the detector-frame distribution of the parameters of the GWs population \citep{Taylor2012, Ezquiaga2022,Karathanasis2022}. Finally, the ``cross-correlation method'' relies on the spatial clustering between GW sources and galaxies surveys \citep{Oguri2016,Mukherjee2020a,Diaz2021}. We expect that lensing systematic bias might also become relevant for future dark sirens surveys, which will consist of many order of magnitudes more events than those considered here. Such black hole events can be detected at distances further away than the NS population, increasing the probability of high magnification. We leave the modelling of lensing bias of dark sirens population to future works. 

Finally, we note that we focused only on the inference of the cosmological parameters. Future works might improve our analysis by inferring jointly cosmological and BNS population parameters. We hypothesize that the inclusion of the latter ones could decrease the amount of bias obtained in our analysis. 

\section{Conclusions}
\label{sec:Conclusion}
We have investigated the impact of lensing in the estimation of the cosmological parameters, using GWs emitted by merging BNS as standard sirens. Differently from previous works in the literature, we treated lensing also as a systematic error instead as only an additional source of statistical error. In the geometric optics approximation, lensing induces a variation in the amplitude of the GW, which translates in a biased estimation of the luminosity distance of the source. We accounted for selection effects due to the modification of SNR of the events and for the non-Gaussian shape of the magnification PDF. 

For our framework, we assumed that 3G ground-base interferometers, in particular ET, together with EM detectors, will be able to collect in a few years of operational time a substantial number of bright sirens to allow for statistical studies. This will open the possibility of testing cosmology using gravitational waves with a precision comparable or even better than the one currently available from EM probes. 

Using the magnification PDF obtained from N-body simulations of \citet{Hadzhiyska2023}, we randomly assigned a value of $\mu$ to each bright siren event in our mock catalog. We allowed also for a tail at high magnifications ($\mu \gtrsim 2$), which could be due, for example, by unidentified strongly lensed events. We included the magnification in the calculation of the SNR and of the biased luminosity distance. We performed our analysis using Fisher forecasts, extended to account for systematic errors. We checked our results with a full likelihood analysis. 

Our fiducial scenario consists of a sample of $N_{\rm events}=3,000$, with an error on the luminosity distance, measured from the GW waveform, given by the sum of the contributions of instrumental uncertainty, weak lensing, peculiar velocity of the source and redshift measurement, assumed to be made spectroscopically. The instrumental uncertainty on the $d_{\rm L}$ is set at ten percent level, supposing that the EM counterpart can help breaking the degeneracy with the angles in the GW amplitude \citep{deSouza2023}. We also use a selection function for EM observation that cuts our redshift distribution at a maximum redshift of $z=2$. For this scenario, we found a bias on the cosmological parameters $H_0$ and $\Omega_{\rm m}$ that fluctuates as a function of the specific realisation of the mock catalog, as can be seen in Figs. \ref{fig:BiasH0real} and \ref{fig:BiasOmreal_marg}, ranging from significance of departure from zero below $1\sigma$, up to $1.5\sigma$ for $H_0$, and $2.5\sigma$ for $\Omega_{\rm m}$. Clearly, if we assumed that $\Omega_{\rm m}$ is perfectly known, the statistical error on $H_0$ would decrease and the impact of the bias would worsen slightly. We also opened the parameter space to the dark energy equation of state parameter $w_{\rm DE}$ in Fig.~\ref{fig:BiasH0_Om_wde}. These results showed that a lensing bias can be expected from future GW data, but its impact on the cosmological parameters strongly depends on the real sample of events that will be collected. When more statistics are considered, i.e. $N_{\rm events}=30,000$, we found that the value of the systematic bias is more stable. 

The assumptions adopted in our analysis play a crucial role in the determination of whether lensing will really be a worrisome source of systematic errors in the 3G epoch. Important aspects are the precision on the luminosity distance measurements and the number of events that we will be able to collect. We found that the larger the error bars on the cosmological parameters or the smaller the number of events, the less significant the systematic bias, as we reported in subsection \ref{subsec:variations}. While we compared the results obtained with different assumptions, future works might implement a complete forecast for the GW parameters estimation analysis and modelling of the EM emission. 

Another point that may impact the analysis is the modelling of the lensing PDF, in particular its high magnification tail. To show this we employed a simple log-normal model for the PDF, whose shape has differences both around the peak and the tail of the distribution. For a more optimistic scenario, we found that a smaller bias is introduced. We also checked that a similar effect is obtained by cutting the high $\mu$ events drawn from our base PDF. Other aspects that can affect the results are the inclination angle and the EM selection function. The former can be restricted to be beamed around zero in order to detect the GRB event. Such events are louder in SNR, thus less affected by lensing selection effects, and in turn, by smaller systematic bias. The latter introduces a selection of the farther events from which the redshift of the source can be determined. Farther events are more likely to undergo high magnifications and spoil the cosmological parameters estimation.

We also presented a simple procedure to try to alleviate the lensing bias. This procedure, even if it is not the final solution, can be regarded as a good and cheap starting point. As can be seen from Fig. \ref{fig:BiasH0real_mitig}, the level of correction that can be achieved depends on the specific realisation and scenario of the mock catalogue, and in the worst case, leave the results unchanged. A safer but less desired solution is to blow up the statistical error bars to make the systematic displacement irrelevant.

Overall, our work highlights the necessity of properly including lensing in the estimation of cosmological parameters from GW, not only as an additional source of noise but also as a possible systematic. On the other hand, we find that the level of accuracy needed in the estimation of the cosmological parameters to appreciate this effect is very high, i.e. of the order of sub-percent level for $H_0$. Given that the current tension in the values of $H_0$ obtained with different probes is of the order of several per-cent, we can safely conclude that BNSs will still be able to give an independent contribution towards overcoming the Hubble tension. On the other hand, a future era of cosmological measurements may open the necessity of reaching such high precision, for example, to probe the nature of dark energy. Although future works are needed in order to better address the impact of lensing in alternative theories of gravity, it is already possible to hypothesize that degeneracy between lensing bias and dark energy effects can arise, making the necessity of correctly modelling lensing very compelling.

\section*{Acknowledgements}

We thank Boet Hoitink for helpful discussions during his thesis project. We are grateful to Tessa Baker, Tanja Hinderer, Macarena Lagos and Otto Hannuksela for helpful feedback on the manuscript. Finally, we thank Jeger Broxterman, Boryana Hadzhiyska, Jia Liu and Ken Osato for help with the weak lensing simulations.
This research was funded in whole, or in part, by the Dutch Research Council (NWO) 24.001.027. For the purpose of open access, a CC BY public copyright license is applied to any Author Accepted Manuscript version arising from this submission.
%%%%%%%%%%%%%%%%%%%%%%%%%%%%%%%%%%%%%%%%%%%%%%%%%%
\section*{Data Availability}

The files containing the mock catalogue of events used in this paper will be made publicly available at the following link: \href{https://github.com/Canev001/GWlensing}{Github/GWlensing}.



\bibliographystyle{mnras}
\bibliography{main} % if your bibtex file is called example.bib

%\begin{thebibliography}{99}
%\end{thebibliography}

%%%%%%%%%%%%%%%%%%%%%%%%%%%%%%%%%%%%%%%%%%%%%%%%%%

%%%%%%%%%%%%%%%%% APPENDICES %%%%%%%%%%%%%%%%%%%%%

\appendix

\section{Additional variation from the fiducial scenario}
\label{AppendixA}
\begin{figure}
    \centering    \includegraphics[width=0.4\textwidth]{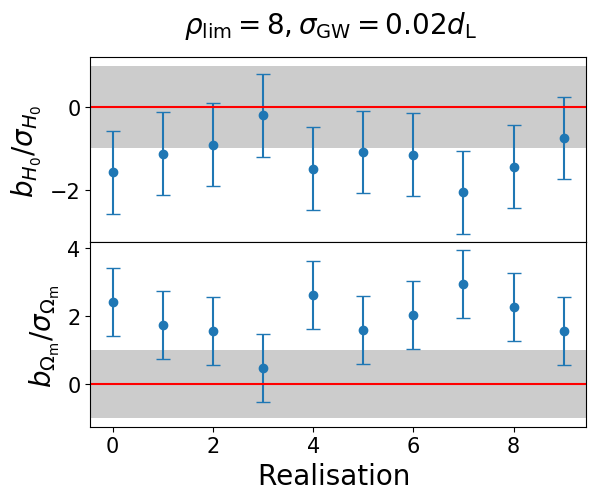}
     \includegraphics[width=0.4\textwidth]{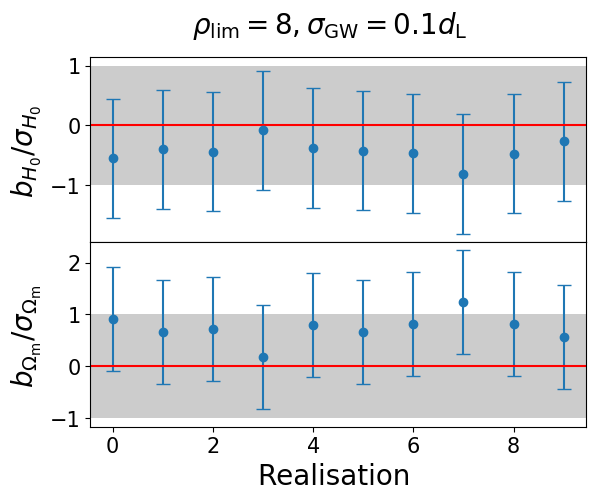}
    \caption{Ratio of the bias and statistical uncertainty on $H_0$ and $\Omega_{\rm m}$ for 10 different realisations of the mock catalog. We vary our fiducial scenario by assuming  $\sigma_{\rm GW}=0.1d_{\rm L}$ with $\rho_{\rm lim}=8$ (upper plot) and $\rho_{\rm lim}=12$ (lower plot).}
    \label{fig:0.1uncertainty}
\end{figure}

In Subsec.~\ref{subsec:variations}, we discuss possible variations of our fiducial scenario that can affect the amount of bias on the cosmological parameters. In this Appendix, we focus more on the influence of the luminosity distance instrumental uncertainty $\sigma_{\rm GW}$ and the limiting SNR $\rho_{\rm lim}$. In our fiducial scenario we consider $\sigma_{\rm GW}=0.1 d_{\rm L}$ and $\rho_{\rm lim}=12$. We also discuss a more optimistic scenario in which $\sigma_{\rm GW}=0.02 d_{\rm L}$ and $\rho_{\rm lim}=8$, assuming that bright standard sirens will be a well-measured small fraction of the total number of events collected by ET. This is in view of the fact that the EM counterpart can help breaking the degeneracy of the luminosity distance with the angles in the amplitude, especially $\iota$ \citep{Chen2019, deSouza2023}. In this case, we find that the statistical error decreases to $0.16\%$ and $1.12\%$ respectively for $H_0$ and $\Omega_{\rm m}$, while the average ratios becomes $b_{H_0}/\sigma_{H_0}=-1$ and $b_{\Omega_{\rm m}}/\sigma_{\Omega_{\rm m}}=2$, as shown in the upper plot of Fig.~\ref{fig:0.1uncertainty}. Thus, in this scenario, although we are assuming to measure the luminosity distance with higher accuracy, the values of the biases remain similar to our fiducial scenario. This result stresses the importance of the limiting SNR threshold of the detector. In our fiducial scenario, we assume $\rho_{\rm lim}=12$ obtained as the sum over the three detectors that compose the triangular configurations of ET. As highlighted in Subsec.~\ref{subsec: add}, increasing $\rho_{\rm lim}$ tends to worsen the bias, since the lensing-induced selection effect becomes more severe. In the lower plot of Fig.~\ref{fig:0.1uncertainty}, we show the results obtained assuming $\rho_{\rm lim}=8$ and  $\sigma_{\rm GW}=0.1d_{\rm L}$. In this case, we find that the average ratios decrease to $b_{H_0}/\sigma_{H_0}=-0.4$ and $b_{\Omega_{\rm m}}/\sigma_{\Omega_{\rm m}}=0.7$. Thus, in this scenario we obtain that the significance of the bias is below $1\sigma$ for both $H_0$ and $\Omega_{\rm m}$, making its contribution on the inferred value of the cosmological parameters negligible.

%%%%%%%%%%%%%%%%%%%%%%%%%%%%%%%%%%%%%%%%%%%%%%%%%%

% Don't change these lines
\bsp	% typesetting comment
\label{lastpage}
\end{document}